\documentclass[apj,twocolumn, appendixfloats, tighten]{openjournal}

\usepackage{lipsum}
\usepackage{newtxtext,newtxmath,enumitem,multirow}

\usepackage[T1]{fontenc}

\DeclareRobustCommand{\VAN}[3]{#2}
\let\VANthebibliography\thebibliography
\def\thebibliography{\DeclareRobustCommand{\VAN}[3]{##3}\VANthebibliography}

\usepackage{xcolor}
\usepackage{textgreek}
\usepackage[utf8]{inputenc}
\usepackage[english]{babel}

\usepackage{hyperref}
\hypersetup{
    unicode, 
    colorlinks=true,
    linkcolor=blue,
    citecolor=blue,
    filecolor=blue,
    urlcolor=blue,
}
\usepackage{color,colortbl}
\definecolor{linkcolor}{rgb}{0.0,0.3,0.5}
\usepackage{tensind}
\tensordelimiter{?}
\DeclareGraphicsExtensions{.bmp,.png,.jpg,.pdf}
\usepackage{verbatim}
\usepackage[normalem]{ulem}
\usepackage{orcidlink}
\usepackage{soul}
\usepackage{graphicx}	
\usepackage{amsmath}	
\usepackage{amssymb}	
\usepackage{wasysym}    
\usepackage{rotating}
\usepackage{xspace}
\usepackage{ulem}
\makeatletter 
  \patchcmd{\NAT@citex}
    {\@citea\NAT@hyper@{%
      \NAT@nmfmt{\NAT@nm}%
      \hyper@natlinkbreak{\NAT@aysep\NAT@spacechar}{\@citeb\@extra@b@citeb}%
      \NAT@date}}
    {\@citea\NAT@nmfmt{\NAT@nm}%
    \NAT@aysep\NAT@spacechar\NAT@hyper@{\NAT@date}}{}{}

  \patchcmd{\NAT@citex}
    {\@citea\NAT@hyper@{%
      \NAT@nmfmt{\NAT@nm}%
      \hyper@natlinkbreak{\NAT@spacechar\NAT@@open\if*#1*\else#1\NAT@spacechar\fi}%
        {\@citeb\@extra@b@citeb}%
      \NAT@date}}
    {\@citea\NAT@nmfmt{\NAT@nm}%
    \NAT@spacechar\NAT@@open\if*#1*\else#1\NAT@spacechar\fi\NAT@hyper@{\NAT@date}}
    {}{}
\makeatother

\newcommand\Msun{\text{M}_{\astrosun}} 
\newcommand\Zsun{\text{Z}_{\astrosun}} 

\newcommand{\HM}{\ion{H}{$_2$}\xspace}
\newcommand{\HI}{\ion{H}{I}\xspace}
\newcommand{\HII}{\ion{H}{II}\xspace}
\newcommand{\HeI}{\ion{He}{I}\xspace}
\newcommand{\HeII}{\ion{He}{II}\xspace}
\newcommand{\HeIII}{\ion{He}{III}\xspace}

\newcommand{\arepo}{{\sc arepo}\xspace}
\newcommand{\areport}{{\sc arepo-rt}\xspace}
\newcommand{\eg}{e.g.\xspace}
\newcommand{\ie}{i.e.\xspace}
\newcommand{\thesan}{\textsc{thesan}\xspace}
\newcommand{\jwst}{\textit{JWST}\xspace}

\newcommand{\hst}{\texttt{HST}\xspace}

\newcommand{\thzoom}{\mbox{\textsc{thesan-zoom}}\xspace}
\newcommand{\thesanone}{\mbox{\textsc{thesan-1}}\xspace}
\newcommand{\thesandarkone}{\mbox{\textsc{thesan-dark-1}}\xspace}
\newcommand{\dd}{\mathrm{d}}
\newcommand{\target}{\textit{target}\xspace}

\newcommand{\centrals}{\textit{centrals}\xspace}
\newcommand{\central}{\textit{central}\xspace}
\newcommand{\satellites}{\textit{satellites}\xspace}
\newcommand{\satellite}{\textit{satellite}\xspace}
\newcommand{\fx}{{``4x''}\xspace}
\newcommand{\ex}{{``8x''}\xspace}
\newcommand{\sx}{{``16x''}\xspace}
\newcommand{\lya}{Ly$\alpha$\xspace}

\shorttitle{Introducing the \thzoom project}
\shortauthors{Kannan et al.}

\begin{document}
\title{Introducing the thesan-zoom project: radiation-hydrodynamic simulations of high-redshift galaxies with a multi-phase interstellar medium\vspace{-1.5cm}}

\author{Rahul Kannan $\orcidlink{0000-0001-6092-2187}$,$^{1\star}$
Ewald Puchwein $\orcidlink{0000-0001-8778-7587}$,$^{2}$
Aaron Smith $\orcidlink{0000-0002-2838-9033}$,$^{3}$
Josh Borrow  $\orcidlink{0000-0002-1327-1921}$,$^{4}$
Enrico Garaldi $\orcidlink{0000-0002-6021-7020}$,$^{5,6}$
Laura Keating $\orcidlink{0000-0001-5211-1958}$,$^{7}$
Mark Vogelsberger $\orcidlink{0000-0001-8593-7692}$,$^8$
Oliver Zier $\orcidlink{0000-0003-1811-8915}$,$^{9,8}$
William McClymont $\orcidlink{0009-0009-5565-3790}$,$^{10,11}$
Xuejian Shen $\orcidlink{0000-0002-6196-823X}$,$^8$
Filip Popovic $\orcidlink{0009-0006-8856-918X}$,$^1$
Sandro Tacchella $\orcidlink{0000-0002-8224-4505}$,$^{10,11}$
Lars Hernquist,$^{9}$
and Volker Springel $\orcidlink{0000-0001-5976-4599}$$^{12}$} 
\thanks{$^\star$ Email: \href{mailto:kannanr@yorku.ca}{kannanr@yorku.ca}}

\affiliation{$^1$ Department of Physics and Astronomy, York University, 4700 Keele Street, Toronto, ON M3J 1P3, Canada \\
$^2$ Leibniz-Institut f\"ur Astrophysik Potsdam, An der Sternwarte 16, 14482 Potsdam, Germany \\
$^3$ Department of Physics, The University of Texas at Dallas, Richardson, TX 75080, USA \\
$^4$ Department of Physics and Astronomy, University of Pennsylvania, 209 South 33rd Street, Philadelphia, PA 19104, USA \\
$^5$ Kavli IPMU (WPI), UTIAS, The University of Tokyo, Kashiwa, Chiba 277-8583, Japan \\
$^6$ Institute for Fundamental Physics of the Universe, via Beirut 2, 34151 Trieste, Italy \\
$^7$ Institute for Astronomy, University of Edinburgh, Blackford Hill, Edinburgh, EH9 3HJ, UK \\
$^8$ Department of Physics, Kavli Institute for Astrophysics and Space Research, Massachusetts Institute of Technology, Cambridge, MA 02139, USA \\
$^9$ Center for Astrophysics $|$ Harvard $\&$ Smithsonian, 60 Garden Street, Cambridge, MA 02138, USA \\
$^{10}$ Kavli Institute for Cosmology, University of Cambridge, Madingley Road, Cambridge CB3 0HA, UK \\
$^{11}$Cavendish Laboratory, University of Cambridge, 19 JJ Thomson Avenue, Cambridge CB3 0HE, UK\\
$^{12}$Max Planck Institute for Astrophysics, Karl-Schwarzschild-Str. 1, D-85741 Garching, Germany}

\begin{abstract}
 We introduce the \thzoom project, a comprehensive suite of high-resolution zoom-in simulations of $14$ high-redshift ($z>3$) galaxies selected from the \thesan simulation volume. This sample encompasses a diverse range of halo masses, with $M_\mathrm{halo} \approx 10^8 - 10^{13}~\Msun$ at $z=3$. At the highest-resolution, the simulations achieve a baryonic mass of $142~\Msun$ and a gravitational softening length of $17~\mathrm{cpc}$. We employ a state-of-the-art multi-phase interstellar medium (ISM) model that self-consistently includes stellar feedback, radiation fields, dust physics, and low-temperature cooling through a non-equilibrium thermochemical network. Our unique framework incorporates the impact of patchy reionization by adopting the large-scale radiation field topology from the parent \thesan simulation box rather than assuming a spatially uniform UV background. In total, \thzoom comprises $60$ simulations, including both fiducial runs and complementary variations designed to investigate the impact of numerical and physical parameters on galaxy properties. The fiducial simulation set reproduces a wealth of high-redshift observational data such as the stellar-to-halo-mass relation, the star-forming main sequence, the Kennicutt--Schmidt relation, and the mass--metallicity relation. While our simulations slightly overestimate the abundance of low-mass and low-luminosity galaxies they agree well with observed stellar and UV luminosity functions at the higher mass end. Moreover, the star-formation rate density closely matches the observational estimates from $z=3-14$. These results indicate that the simulations effectively reproduce many of the essential characteristics of high-redshift galaxies, providing a realistic framework to interpret the exciting new observations from \jwst.
\end{abstract}


\maketitle

\section{Introduction}
\label{sec:intro}

The study of primitive stars and galaxies is an exciting new frontier in astrophysics and cosmology. These early structures form within a few hundred million years after the Big Bang through the gravitational amplification of primordial density fluctuations, which collapse into dark matter halos \citep{Zeldovich1970}. Gas flows into these halos, cools radiatively, and forms the first stars and galaxies, marking the beginning of Cosmic Dawn (CD). The intense high-energy radiation from these stars dramatically alters their surroundings by transforming the previously cold, neutral gas into a hot, ionized medium \citep{Barkana2001}. This process, called reionization, is initially patchy, with ionized bubbles surrounding the most energetic sources.  These bubbles grow bigger and more numerous and eventually overlap as the ionizing radiation output from galaxies increases, ionizing and heating nearly all the low-density gas in the Universe. Understanding these early epochs of CD and reionization (EoR) is necessary because they form an important evolutionary link between the smooth matter distribution at early times and the complex structures observed today. 

With the launch of the James Webb Space Telescope (\jwst), we have entered an exciting new high-precision era in the study of high-redshift structure formation and reionization.  The large $6.5 \, \mathrm{m}$ mirror and unprecedented near- to mid-infrared instrumentation have allowed us to detect the rest-frame optical emission \citep{Kalirai2018, Williams2018} of galaxies within the first billion years after the Big Bang. These early observations reveal a surprisingly large abundance of incredibly bright, massive galaxies at $z>8$ \citep{Labbe2023, Finkelstein2023, Weibel2024}, which seem to be in tension with several semi-analytical models \citep{Behroozi2020, MH2023, Yung2024} and numerical simulations \citep{Vijayan2021, KannanThesan, Kannan2023}.  While some discrepancies may stem from measurement uncertainties, such as photometric redshift misestimation \citep{Naidu2022, Zavala2023} or emission line contamination in broadband filters \citep{Endsley2023, Tacchella2023}, there is growing evidence \citep{Willot2023, Castellano2024} that the population of high-redshift galaxies may differ significantly from what many models of galaxy formation predict. 

Simulations such as \textsc{firstlight}, \citep{Ceverino2017} \textsc{fire-2} \citep{Hopkins2018}, and \textsc{serra} \citep{Serra2022} suggest that large gas fractions and surface densities can lead to high star formation efficiencies (SFE) in high-redshift galaxies  \citep{Bassini2023}. Moreover, the significant reduction of mechanical energy in stellar winds from metal-poor stars  \citep{Dekel2023} can trigger feedback-free starbursts (FFBs) in the metal-deficient primitive Universe. However, recent studies have shown that the galaxy-averaged star formation efficiency can remain relatively low even if the local efficiency in dense gas nears unity, due to a highly inhomogeneous gas density distribution with most of the gas at a low surface density \citep{Sun2025}. Another plausible explanation could be a significant increase in the stochasticity of star formation \citep{Shen2023, Sun2023}. Due to the steep faint-end slope, more galaxies will scatter from faint to bright luminosities than from bright to faint, which can lead to a shallower bright-end slope. Alternatively, non-standard cosmological phenomena like early dark energy \citep{EDE2022, Shen2024}, primordial black holes \citep{Liu2022} or cosmic string loops \citep{Jiao2023, Koehler2024} can potentially seed structures in the early Universe leading to a larger than expected abundance of massive galaxies at high redshifts. 

Observations suggest that these early galaxies are predominantly metal-poor \citep{Curti2023} and relatively dust-free \citep{Bakx2023}, and a large majority of them have high specific star-formation rates and bursty star formation histories \citep{Endsley2023}, especially at the low-mass end. The spectral energy distributions (SEDs) are often dominated by very young stellar populations, producing strong emission lines with high equivalent widths \citep{Rinaldi2023, Withers2023}. Moreover, they have emission line ratios that indicate higher degrees of ionization and lower metal abundances \citep{Sanders2023} than their low-redshift counterparts. The relatively young and low metallicity stars have very high production efficiency of ionizing photons with the median observed $\xi_\mathrm{ion} \gtrsim 10^{25.5} \, \mathrm{erg}^{-1} \, \mathrm{Hz}$, suggesting they could play a significant role in reionizing the Universe \citep{Simmonds2023}. Remarkably, \jwst detections of damping wing absorption redward of the \lya line \citep{Hsiao2023, Umeda2023} hint at ionized bubbles that may be substantially larger than predicted in many analytic or simulation-based models \citep{Neyer2023}, although alternative explanations involve higher \lya luminosities or smaller bubbles \citep{Keating2023}.

\jwst has revealed additional surprises, including the existence of disk-dominated galaxies just $600 \, \mathrm{Myr}$ after the Big Bang \citep{Li2023, Nelson2023, Robertson2023}. Although stable disks are common in the low-redshift Universe, the high merger and accretion rates are not conducive for disk formation \citep[e.g.][]{Semenov2024} at such early times. There have also been a few observations of non-star forming (quenched) low-mass galaxies in the reionization epoch \citep{Looser2023, Strait2023}. It is unclear what caused these galaxies to quench, but there is some evidence that the combined effect of multiple supernovae after a starburst event can drive large-scale outflows that temporarily quench a galaxy \citep{Dome2024}. Intriguingly, there have also been observations of extremely metal-poor stars \citep{Vanzella2023}, and even the discovery of a bound proto-globular cluster, just $460 \, \mathrm{Myr}$ after the Big Bang \citep{Adamo2024}.

To interpret these new high-redshift observations, theoretical models must capture both the primordial galaxy formation process and their impact on the environment. This requires accurately modeling the multitude of physical processes that sculpt the properties of galaxies \citep[see ][ for a review]{Vogelsberger2020} coupled to accurate radiation hydrodynamics (RHD). The complexity and numerical cost of including on-the-fly radiative transfer has forced almost all of the state-of-the-art galaxy formation simulations to approximate the radiation field as either a spatially uniform background \citep[see for example, ][]{Hopkins2014, KannanMagicc, Vogelsberger2014, Schaye2015, Schaye2023, Wang2015, Springel2018, Dave2019, Pakmor2023} or use approximations like optically thin transport \citep{Kannan2014, Kannan2016, Hopkins2018, Obreja2019, Feldmann2023, Waterval2024, Zhu2024}. However, recent advancements in computational power and algorithmic improvements have allowed for the inclusion of on-the-fly radiative transfer in large-scale galaxy formation simulations \citep{Gnedin2001, Rosdahl2015, Kannan2019}. Many of these coupled galaxy formation and RHD simulations prioritize modeling the reionization epoch due to the considerable influence of radiation fields on galaxies and their environment in this era.

Simulating large representative volumes ($\gtrsim 100 \, \mathrm{cMpc}$) of the Universe is necessary to capture the large-scale statistical properties of the reionization process and high-resolution ($\sim 100 \, \mathrm{pc}$) is required to predict the resolved properties of the galaxies responsible for it. Such large-scale simulations have recently made progress in understanding the reionization epoch by making predictions for the high-redshift UV luminosity function \citep{Ocvirk2020, KannanThesan}, impact of photo-heating feedback during reionization in low-mass galaxies \citep{Gnedin2014, Pawlik2017, Borrow2023},  size distribution of ionized bubbles \citep{Neyer2023, Thelie2023}, morphology of 21 cm emission \citep{Kaurov2016, KannanThesan}, mean free path of ionizing photons \citep{Garaldi2022, Lewis2022} and Ly-$\alpha$ emission and transmission during the EoR \citep{Gronke2021, Park2021, Smith2022}. These simulations have also been used to provide insights into some resolved properties of galaxies, like the fraction of ionizing photons that escape the galaxy \citep{Lewis2020, Yeh2023} and various emission line luminosities arising from highly ionized \ion{H}{II} regions \citep{Kannan2022, Hirschmann2023}. However, the resolution and/or simplistic modeling of the multi-phase interstellar medium (ISM) reduce the accuracy of these predictions. This is because the escape of LyC photons heavily depends on the structure of gas on small scales ($\lesssim 1~\rm{pc}$) around the star \citep{Smith2022b}, which is not properly modeled in these large-volume simulations. Additionally, most of these simulations rely on artificially painting \ion{H}{II} regions around newly formed stars to predict emission line luminosities. The models of these \ion{H}{II} regions often use predetermined values for densities and ionization parameters that may not accurately reflect the actual ISM conditions, especially in high-redshift environments.

Alternatively, one can choose to put most of the computational effort into modeling the resolved properties of a small number of high-redshift galaxies \citep{ Rosdahl2018, Hopkins2020, Serra2022, Bhagwat2023,Hopkins2024}, sacrificing volume for resolution. This, in principle, allows for better modeling of processes like gas cooling in the ISM, photoionization, photoheating, and stellar feedback. These simulations have been used to gain insights into how the escape fraction of ionizing photons scales with other galaxy properties \citep{Rosdahl2022}, the role of AGN feedback in the reionization process \citep{Trebitsch2021} and modeling the emission line luminosities of high-redshift galaxies \citep{Serra2}. Unfortunately, most of these simulations employ galaxy formation models that are not sufficiently well tested  \citep{Xu2016, Ceverino2017, Rosdahl2022} because they have been specifically designed and evaluated in simulations that mainly predict observables above $z \gtrsim 6$. When these galaxy formation models are used to simulate galaxies down to lower redshifts ($z\lesssim3$), they produce an order of magnitude more stars than what is expected and bulge-dominated galaxies with centrally peaked rotation curves \citep{Mitchell2021}. Moreover, many of them lack one or more of the major ingredients that regulate the properties of high-redshift galaxies like, dust content, molecular hydrogen chemistry, and/or a realistic external radiation field that is spatially and temporally varying. 

In this paper, we introduce a novel simulation campaign designed to provide a realistic simulation counterpart to the extensive array of \jwst observations of high-redshift galaxies. We select $14$ high-redshift galaxies, from the large-volume coupled galaxy formation and reionization simulation, \thesanone \citep{KannanThesan, Smith2022, Garaldi2022}, and re-simulate them at much higher resolution using the zoom-in technique. We use a significantly evolved version of the \textsc{smuggle} multi-phase ISM model that self-consistently accounts for the effects of supernova feedback, radiation fields, dust physics, and low temperature cooling via molecular hydrogen \citep{Marinacci2019, Kannan2020a, Kannan2021}. The highest-resolution simulations in our suite reach a baryonic mass and spatial resolution of $142 \, \Msun$ and $\sim 17 \, \mathrm{cpc}$, respectively. The simulations also follow the propagation of radiation across the boundary of the high-resolution region to ensure incoming radiation flows present in the \thesanone simulation are seen by the high-resolution region of the zoom-in run. This allows for a more realistic treatment of reionization for objects that are inefficient sources of ionizing radiation and reionize outside-in. We also perform a set of supplementary simulations designed to investigate the effect of numerical resolution and various physical parameters on galaxy properties. In total, this gives us a sample of 60 simulations, which are used to investigate the properties of high-redshift galaxies and interpret the exciting new observations from \jwst. The methodology is introduced in Section~\ref{sec:methods} with the main results presented in Section~\ref{sec:results} and the conclusions outlined in Section~\ref{sec:conclusions}.

\section{Methods}
\label{sec:methods}
The simulations use \areport \citep{Kannan2019}, which is a radiation hydrodynamic extension to the moving mesh hydrodynamics code \arepo \citep{Springel2010,Weinberger2020}. The radiation hydrodynamic equations are solved using a finite-volume approach on an unstructured mesh. This mesh is built from the Voronoi tessellation of a set of mesh-generating points, which are generally free to follow the motion of gas with the restriction that the mesh is frequently regularized according to the algorithm described in \cite{Vogelsberger2012}. A quasi-Lagrangian solution to the hydrodynamic equations is achieved by solving them at interfaces between moving mesh cells in the rest frame of the interface. Second-order convergence is achieved using a piecewise linear approach, which calculates gradients using a least square fit (LSF) estimate that performs well even on highly distorted meshes \citep{Pakmor2016}. Time integration is performed using Heun’s method, which computes the fluxes as an average of fluxes at the beginning and end of the time step. The mesh geometry of the second half of the current time step is used for the first half of the next time step, essentially requiring only one mesh construction per step. The motion of the mesh-generating points naturally adjusts the mesh resolution to the underlying density field and is, therefore, well suited to simulate systems with a large dynamical range.

The gravitational forces are computed using a hybrid approach which solves the short-range forces using a hierarchical oct-tree algorithm \citep{Barnes1986}, while the long-range forces are calculated using the particle mesh method \citep{Aarseth2003}. To accelerate the calculations of the gravitational force, we employ hierarchical time integration as described in \cite{Gadget4}. This method enables us to build the tree solely for the currently active particles, which is particularly advantageous in high-resolution simulations characterized by a deep time-bin hierarchy. We use a simple but effective method to deal with the correlated force errors that occur at large node boundaries in cosmological simulations due to almost static particle distributions at high redshift. This involves randomizing the relative locations of the particle set with respect to the computational box each time a new domain decomposition is computed \citep{Gadget4}.

\subsection{Radiation transport}
Radiation fields are simulated by casting the radiative transfer equation into a set of hyperbolic conservation laws for photon number density ($N_i$; where 'i' denotes the frequency bin) and photon flux (${\bf{F}}_i$) by taking its zeroth and first moments, respectively \citep{Kannan2019}. Since this work presents cosmological simulations, we evolve the comoving photon number density ($\tilde{N}_i$) and photon flux (${\bf{\tilde{F}}}_i$), which are defined as $\tilde{N}_i = a^3 N_i$ and ${\bf{\tilde{F}}}_i = a^3 {\bf{F}}_i$. The cube of the scale factor ($a$) is multiplied to the physical quantities to account for the loss of photon energy due to cosmological expansion. Under the approximation that the Universe does not expand significantly before a photon is absorbed \citep{Gnedin2001}, the transport equations take the form \citep{Wu2019a, KannanThesan}
\begin{equation}
     \frac{\partial \tilde{N}_i}{\partial t} + \frac{1}{a} \nabla \cdot \tilde{\bf{F}}_i = 0 \, ,
     \end{equation}
     \begin{equation}
     \frac{\partial {{\tilde{\bf{F}}}}_i} {\partial t} + \frac{\tilde{c}^2}{a} \nabla \cdot {\tilde{\mathbb{P}}_i} = 0 \,, \\
\end{equation}
where $\tilde{c}$ is the reduced speed of light, which for our runs is set to $\tilde{c} = 0.01\,c$\footnote{We note that this value is much lower than the one used in the \thesan simulations ($0.2c$). High light speeds were necessary because the ionization-fronts at the tail end of reionization can reach speeds as high as $0.1\,c$ as they sweep through the low-density voids \citep{Daloisio2019}. However, for galaxy formation simulations, which mainly focus on high-density regions in and around halos, light speeds as low as $1000~\rm{km}~\rm{s}^{-1}$ are sufficient to achieve converged results while significantly relaxing the stringent time-stepping criteria for radiation transport \citep{Kannan2020a}.} (where $c$ is the speed of light in vacuum),
and $\tilde{\mathbb{P}}_i$ is the comoving pressure tensor, which is related to the photon number density by the Eddington tensor. 

These equations are closed using the Eddington tensor formalism, with the specific form of the Eddington tensor given by the M1 closure approximation \citep{Levermore1984, Dubroca1999}. Second-order accuracy is achieved by using a slope-limited piecewise linear extrapolation using the same local LSF gradient estimate used to solve the hydrodynamic equations. The algorithm is fully conservative and compatible with the individual time-stepping scheme of \arepo. The high speed of light demands very small timesteps, forcing other computationally expensive parts of the code, such as mesh construction and gravity force calculations, to be called more often than actually required. To overcome this issue, for each hydro timestep, the radiation transport is sub-cycled sixteen times according to the algorithm described in Appendix A of \citet{Kannan2019}. Finally, we adopt a novel node-to-node communication strategy that utilizes shared memory to substitute intra-node communication with direct memory access. Combining multiple inter-node messages into a single message substantially enhances network bandwidth utilization and performance for large-scale simulations like the ones presented in this work \citep{Zier2024}.

\subsection{The thermochemical network and gas cooling}
\label{sec:cooling}
The radiation fields are coupled to the gas via a non-equilibrium thermochemical network,  comprehensively outlined in Section 2.1 of \citet{Kannan2020b}. Briefly, it models primordial chemistry and cooling ($\Lambda_\mathrm{p}$), uses tabulated cooling rates for metals ($\Lambda_\mathrm{M}$), photoelectric heating ($\Lambda_\mathrm{PE}$), cooling from dust-gas-radiation field ($\Lambda_\mathrm{D}$) interactions and Compton cooling off the cosmic microwave background (CMB; $\Lambda_\mathrm{C}$). The total cooling ($\Lambda_\mathrm{tot}$) is given by
\begin{equation}
\begin{split}
    \Lambda_\mathrm{tot} &= \Lambda_\mathrm{p}(n_j, N_i, T) + \frac{Z}{\Zsun}\Lambda_\mathrm{M}(T, \rho, z)
     + \Lambda_\mathrm{PE}(D, T, \rho, N_\mathrm{FUV})\\ &+ \Lambda_\mathrm{D}(D, T, \rho, N_\mathrm{IR}) + \Lambda_\mathrm{C}(\rho, T, z) \, ,
\end{split}
\end{equation}
where $n_j$ is the number density of the ionic species '$j$', tracked by the non-equilibrium network, $T$ is the temperature of the gas, $\rho$ is the density, $D$ is the dust-to-gas ratio, $Z$ is the metallicity, $\Zsun$ is the solar metallicity, $z$ is the redshift and $N_\mathrm{FUV}$ and $N_\mathrm{IR}$ are the radiation field intensity in the far-UV and the infrared bands, respectively. 

The primordial thermochemical network calculates non-equilibrium abundances of six chemical species, namely  $ \HM, \HI, \HII, \HeI, \HeII,$ and $\HeIII$. The network describing the thermochemistry of atomic Hydrogen and Helium is based on Equations (49)--(51) in \citet{Kannan2019}. This is supplemented by a simple model for molecular hydrogen that is coupled to the dust formation and destruction model (see Section~\ref{sec:dust} for more details) as outlined in \citet{Kannan2020b}. These abundances are then used to calculate the cooling and heating rates from hydrogen and helium. We note that this model reproduces the right amount of molecular gas and matches the molecular Kennicutt--Schmidt relation in idealized simulations of Milky Way like galaxies.

Metal cooling is implemented assuming ionization equilibrium for a given portion of dust-free and optically thin gas in a UV background radiation field given by \citet{FG09}. The cooling rate is computed from a look-up table containing the pre-calculated cooling values computed from CLOUDY (see \citealt{Vogelsberger2013}, for more details). We note that the metal cooling rate assumes a spatially constant metagalactic UV background and therefore does not account for variations in the radiation field intensities. More explicit modeling requires calculating non-equilibrium metal ionization states coupled to radiation hydrodynamics \citep{Katz2022, Richings2022}. However, we expect this approximation to have only a small impact on our results in the context of the present galaxy formation model. Far-UV photons (FUV; $5.8-11.2 \,  \mathrm{eV}$) have high enough energies to knock off electrons from dust grains, which can, in turn, heat the interstellar medium of galaxies. This heating rate depends on the dust content of the gas and the intensity of the local FUV radiation as outlined in \citet{Wolfire2003}. The final cooling term arises from the dust–gas energy exchange via collisions and is given by \citep{Burke1983}
\begin{equation}
\begin{split}
    \Lambda_\mathrm{D} &= \beta \left(\frac{D}{0.01}\right) \left(\frac{0.1~\mu\mathrm{m}}{a}\right) \sqrt{\frac{T}{1~\mathrm{K}}} \left(\frac{T-T_d}{1~\mathrm{K}}\right) \left(\frac{n_\mathrm{H}}{1 ~\mathrm{cm}^{-3}}\right)^2 \\
    \mathrm{with}~\beta &= 1.356 \times 10^{-33}~\mathrm{erg}~\mathrm{cm}^{-3}~\mathrm{s}^{-1}~,
\end{split}
\end{equation}
where $a$ is the grain size, which we set to $0.1~\mu\mathrm{m}$, $T$ is the gas temperature, $T_\mathrm{d}$ is the dust temperature, and $n_\mathrm{H}$ is the gas density is units of particles per cubic centimeter. The values of $D$ and $T_\mathrm{d}$ are self-consistently modeled by the dust model described in the next section. Finally, we note that we do not include high-energy ionization sources such as cosmic rays and X-rays that influence the electron fraction in the cold and warm neutral phases of the ISM \citep{Bialy2019}, which may lead to an underestimate of the heating rate in these phases, although the impact on high-redshift galaxies is largely unconstrained. We intend to include these processes in the future.

\subsection{Dust physics}
\label{sec:dust}
Dust is modeled as a scalar property of the gas cells, which neglects relative motion between these two components and passively advects dust across gas cells during the hydrodynamical step as outlined in \citet{McKinnon2016, McKinnon2017}. The model tracks the mass of dust in five chemical species (C, O, Mg, Si, and Fe). Dust production during stellar evolution is modeled assuming part of the metals returned to the ISM condensate into dust within a single simulation timestep. The condensation fraction depends on the stellar evolution phase, with asymptotic giant branch (AGB) stars and supernovae (SN) being the major contributors \citep{Dwek1998}.  

Once produced, dust mass is allowed to increase through the deposition of gas-phase metals onto existing dust grains. The local instantaneous dust growth rate is
\begin{equation}
    \frac{\dd{M_{\mathrm{dust}}}}{ \dd{t}} = \left( 1 - \frac{M_{\mathrm{dust}}}{M_{\mathrm{metal}}} \right) \frac{M_{\mathrm{dust}}}{\tau_\mathrm{g}}~, \qquad\qquad \text{(Growth)}
\end{equation}
where $M_\mathrm{dust}$ and $M_{\mathrm{metal}}$ are the mass of dust and gas-phase metals in the cell. $\tau_\mathrm{g}$ is the growth time scale, which is defined as
\begin{equation}
\begin{split}
 \tau_\mathrm{g} &= \tau_\mathrm{g}^\mathrm{ref} \left(\frac{\rho^\mathrm{ref}}{\rho}\right) \sqrt{\frac{T^\mathrm{ref}} {T}} \frac{\Zsun}{Z}, \, \mathrm{if} \, T<300~\mathrm{K}, \\
 &= \infty, \, \mathrm{otherwise} \, .
 \end{split}
 \label{eq:dgr}
\end{equation}
$\tau_\mathrm{g}^\mathrm{ref}$ is the reference growth time which depends on atom–grain collision, sticking efficiencies, and grain cross-sections and is set to $200 \, \mathrm{Myr}$. $\rho^\mathrm{ref}$ and $T^\mathrm{ref}$ are the reference density and temperature, which are set to 1 H atom $\mathrm{cm}^{-3}$ and $20\, \mathrm{K}$, respectively. We note that the additional metallicity dependence and the cutoff temperature are new features added to the dust model presented in \citet{McKinnon2016}. These additional dependencies were introduced to match the sudden boost in dust abundances in gas with close to solar metallicities observed in the local Universe \citep{RR2014}. 

The local instantaneous dust destruction rate by shocks \citep[\eg][]{Seab&Shull83, Seab87, Jones+94} and  sputtering \citep[both thermal and non-thermal, \eg][]{Draine&Salpeter79b, Tielens+94} is given by
\begin{equation}
    \frac{\dd M_{\mathrm{dust}}}{\dd t} = -\frac{M_{\mathrm{dust}}}{\tau_\mathrm{sh}} -\frac{M_{\mathrm{dust}}}{\tau_\mathrm{sp}/3}~. \qquad\qquad \text{(Destruction)}
\end{equation}
$\tau_\mathrm{sh}$ and $\tau_\mathrm{sp}$ are the shock- and sputtering-driven destruction time scales, respectively. SN shocks only destroy dust in gas particles currently experiencing SN feedback, from nearby newly formed stars.  The shock-driven destruction timescale depends upon the local dust mass, the grain-destruction efficiency of SN shocks, and the typical shock velocity \citep[see Eqs. 16 of][]{Kannan2020b}. For computing $\tau_\mathrm{sp}$, we assume that thermal sputtering processes dominate over non-thermal ones. Hence it solely depends upon the local gas temperature, density, and grain size. Finally, we track the temperature of dust grains by modeling the energy exchange between the infra-red (IR, $0.1-1$ eV) radiation field and dust grains
\begin{equation}
    \Lambda_\mathrm{R} = \kappa_\mathrm{P}\rho c (a_r T_\mathrm{d}^4 - E_\mathrm{IR})~,
\end{equation}
where $\kappa_\mathrm{P}$ is the Planck mean opacity, $a_r$ is the radiation constant, and $E_\mathrm{IR}$ is the sum of the energy densities of the radiation field in the infrared frequency bin and the energy density of the cosmic microwave background \citep{Kannan2019}. The dust temperature is calculated by solving the instantaneous equilibrium condition $\Lambda_\mathrm{D} + \Lambda_\mathrm{R} = 0$ using Newton’s method for root-finding. This model reproduces the observed distributions of dust-to-gas ratios and dust temperatures \citep{Utomo2019} in nearby galaxies \citep{Kannan2020b, Kannan2021}.

\begin{table*}
	\centering
	\caption{\textup{Table outlining the frequency discretization of the radiation field used in the \thzoom simulations. It lists the frequency bin name (first column), photon energy range (second column), the mean photoionization cross section ($\sigma$) for the different species ($\ion{H}{2}$, third column; $\ion{H}{I}$, fourth column; $\ion{He}{I}$, fifth column; $\ion{He}{II}$, sixth column), the energy injected into the gas per absorbed photon ($\mathcal{E}$) for the different species ($\ion{H}{2}$, seventh column; $\ion{H}{I}$, eighth column; $\ion{He}{I}$, ninth column; $\ion{He}{II}$, tenth column), the mean radiation pressure ($\mathcal{P}$) for the different species ($\ion{H}{2}$, eleventh column; $\ion{H}{I}$, twelfth column; $\ion{He}{I}$, thirteenth column; $\ion{He}{II}$, fourteenth column), the mean energy per photon ($e$; fifteenth column) and the dust-photon interaction cross section ($\kappa_\mathrm{d}$, sixteenth column). The dust opacities for the Optical, FUV, LW, and ionizing bins (EUV1, EUV2, EUV3) are constant, while the IR opacities are calculated assuming a certain grain size distribution and dust temperature as outlined in Appendix C of \citet{Kannan2019}.}}
	\label{table:rad}
 \small\addtolength{\tabcolsep}{-0.3pt}
	\begin{tabular}{cccccccccccccccc} 
		\hline
        \vspace{-0.25cm}\\
		Bin & Range & $\sigma_\ion{H}{2}$ & $\sigma_\ion{H}{I}$ & $\sigma_\ion{He}{I}$ & $\sigma_\ion{He}{II}$ & $\mathcal{E}_\ion{H}{2}$ & $\mathcal{E}_\ion{H}{I}$ & $\mathcal{E}_\ion{He}{I}$  & $\mathcal{E}_\ion{He}{II}$ & $\mathcal{P}_\ion{H}{2}$ & $\mathcal{P}_\ion{H}{I}$ & $\mathcal{P}_\ion{He}{I}$  & $\mathcal{P}_\ion{He}{II}$ & $e$ & $\kappa_\mathrm{d}$\vspace{0.05cm}\\  
		 & $[$eV$]$& [Mb] & [Mb] & [Mb]  & [Mb]  & [eV] & [eV] & [eV] & [eV] & [eV] & [eV] & [eV] & [eV] & [eV] & [$\mathrm{cm}^2 \mathrm{g}^{-1}$]\vspace{0.05cm}\\
		\hline
        \vspace{-0.2cm}\\
        IR & $0.1-1.0$ & 0 & 0 & 0 & 0 & 0 & 0 & 0 & 0 & 0 & 0 & 0  & 0 & 0.65 & $f(T_\mathrm{d})$ \vspace{0.05cm}\\
        Optical & $1.0 - 5.8$ & 0 & 0 & 0 & 0 & 0 & 0 & 0 & 0 & 0 & 0 & 0 & 0 & 3.68 & 150 \vspace{0.05cm}\\
        FUV & $5.8-11.2$ & 0 & 0 & 0 & 0 & 0 & 0 & 0 & 0 & 0 & 0 & 0 & 0 & 8.47 & 1500\vspace{0.05cm}\\
        LW & $11.2-13.6$ & 0.21 & 0 & 0 & 0 & 0 & 0 & 0 & 0 & 12.26 & 0 & 0 & 0 & 12.36 & 1000 \vspace{0.05cm}\\
		EUV1 & $13.6 - 24.6$ & $5.09$ & $3.31$ & 0 & 0 & $3.94$ & $3.25$ & 0 & 0 & 19.14 & 16.85 & 0 & 0 & 18.16 & 1000 \vspace{0.05cm}\\
		EUV2 & $24.6 - 54.4$ & $2.14$ & $0.70$ & $4.5$ & 0 & 13.96 & 15.64 & 4.74 & 0 & 29.16 & 29.24 & 29.34 & 0 & 32.05 & 1000\vspace{0.05cm}\\
		EUV3 & $54.4 - \infty$ & $0.32$ & $0.11$  & $0.77$ & $1.42$ & 41.23 & 42.87 & 31.87 & 2.10 & 56.46 & 56.46 & 56.47 & 56.50 & 56.99 & 1000\vspace{0.05cm}\\
		\hline
	\end{tabular}
\end{table*}

\subsection{Star formation}
\label{sec:sf}

Star are allowed to form in gas cells that have a thermal Jeans length smaller than the cell size, 
\begin{equation}
L_\mathrm{J} = \sqrt{\pi c_\mathrm{s}^2 / G \rho} < \Delta x~,
\end{equation}
where $c_\mathrm{s}$ is the sound speed, $G$ is Newton's gravitational constant, $\rho$ is the density of the cell, and $\Delta x$ is the cell radius which is calculated assuming spherical geometry. This ensures that the stars are only formed when the Jeans length becomes unresolved and the gravitational collapse can no longer be followed accurately \citep{Truelove1997}. This condition can allow for low-density gas outside the ISM of galaxies to form stars if it is relatively cold. To avoid this issue, we include an additional condition that the gas must be at least denser than $n_\mathrm{H} = 10~\mathrm{cm}^{-3}$ to be eligible to form stars. Gas cells that satisfy both criteria are converted to stars using the usual probabilistic approach \citep{Springel2003}, which assumes that the star-formation rate 
is proportional to the gas density and inversely proportional to the local free fall time ($t_\mathrm{ff} = \sqrt{3\pi/32G \rho}$)
\begin{equation}
    \mathrm{SFR} = \epsilon_\mathrm{ff}\,\rho/t_\mathrm{ff}~,
\end{equation} where $\epsilon_\mathrm{ff}$ is the star formation efficiency per free-fall time of gas. In our simulations, we set this efficiency to $100\%$ to prevent the unresolved gas from artificially collapsing to high densities \citep{Hu2023}. We have performed simulations with other values of $\epsilon_\mathrm{ff}$, varying it all the way from $1-100\%$. This parameter does not have any significant effect on the global properties of the simulated galaxies, which is in agreement with previous works that investigated the impact of $\epsilon_\mathrm{ff}$ in resolved ISM simulations \citep{Hopkins2011, Hopkins2013, Hopkins2018}. The probability of a cell forming a star is
\begin{equation}
    p_\mathrm{star} = 1- \exp\left(-\mathrm{SFR} \frac{\Delta t}{m_\star}\right),
\end{equation}
where $m_\star = \mathrm{min}(m_\mathrm{cell}, m_\star^\mathrm{max})$, $\Delta t$ is the time step of the cell and $m_\star^\mathrm{max}$ is the maximum stellar mass set to $m_\star^\mathrm{max} = 2 m_\mathrm{gas}$ where $m_\mathrm{gas}$ is the baryonic mass resolution. Collisionless particles of mass $m_\star$, representing stellar populations, are formed stochastically from the gas, with the probability of forming one drawn from a uniform distribution.

\subsection{Stellar feedback}
The simulations model three forms of stellar feedback: radiation feedback, stellar winds, and supernova (SN) feedback. We model photoheating, radiation pressure, and photoelectric heating self-consistently using a radiative transfer scheme, with star particles as the source of local radiation.

\subsubsection{Radiative feedback}
The luminosity and spectral energy density of stars is a complex function of age and metallicity taken from the Binary Population and Spectral Synthesis models (BPASS; \citealt{BPASS2017}).  The radiation is injected into $16$ cells closest to the star particle. In addition, the photon flux $\textbf{F}$ is directed radially outwards from the star particle, with magnitude $\textbf{F} = \tilde{c}E$, to ensure that the full radiation pressure force is accounted for even if the cell’s optical depth is larger than one \citep{Kannan2020a}.

The radiation fields are coupled to the gas via the non-equilibrium chemical network that follows the abundance of molecular and atomic hydrogen and helium as outlined in Section~\ref{sec:cooling}. As moment-based methods do not follow individual rays, it is not possible to follow the differing shapes of the radiation spectrum from different age and metallicity sources. Therefore, we cannot capture the cell-to-cell variation of the radiation field. However, important radiation quantities like the mean ionization cross section ($\sigma$), mean photoheating rate ($\mathcal{E}$), mean radiation pressure ($\mathcal{P}$) and mean photon energy ($\epsilon$) in each frequency bin are roughly constant and do not vary significantly with
the metallicity and age of the star \citep{Rosdahl2015}. We, therefore, calculate them using a $2$~Myr spectrum at quarter solar metallicity and employ the same values for all cells throughout the simulation. We discretize the radiation field into seven radiation band, namely, the infra-red \mbox{(IR, $0.1-1$ eV)} band, the optical band \mbox{(Opt, $1-5.8$ eV)}, the far ultraviolet band \mbox{(FUV, $5.8-11.2$ eV)}, the Lyman–Warner band \mbox{(LW, $11.2-13.6$ eV)}, hydrogen ionizing bin \mbox{(EUV1, $13.6-24.6$ eV)}, He I ionizing bin \mbox{(EUV2, $24.6-54.4$ eV)}, and finally the He II ionizing bin \mbox{(EUV3, $54.4-\infty$ eV)}. The mean ionization cross-section, photoheating rate, and mean radiation pressure for each bin are tabulated in Table~\ref{table:rad}. Finally, we include explicit correction factors to account for the over-ionization, overheating, and missing photons issues that arise in unresolved/partially resolved \ion{H}{II} regions as described in \citet{Deng2024a}.

\subsubsection{Stellar winds and supernovae}
Energy and momentum injection from stellar winds and SN explosions are incorporated according to the \textsc{smuggle} \citep{Marinacci2019} model which closely follows the prescriptions outlined in the \textsc{fire-2} model. Briefly, empirical prescriptions outlined in \citet{Hopkins2018} for the mass
loss rate and the energy of the stellar winds are used to calculate the total momentum of the winds. The mass, momentum, and energy from stellar winds are then injected in a continuous manner in the rest frame of the star and then transformed back into the reference frame of the simulations. To be consistent with the metal and dust enrichment routines \citep{Vogelsberger2013, McKinnon2016}, we assume a Chabrier Initial Mass Function \citep[IMF; ][]{Chabrier2003} with minimum and maximum stellar masses set to $0.1~\Msun$ and $100~\Msun$ respectively. All stars with $M_\star \geq 8~\Msun$ are presumed to explode as SN and inject the canonical $10^{51}~\mathrm{ergs}$ into the surrounding ISM. However, injecting just the energy can lead to over-cooling in galaxies \citep{DV2012} since the Sedov--Taylor phase of the SN blast wave is often not resolved in the simulation. To overcome this, the terminal momentum, defined as the momentum that is generated by the $P~\mathrm{d}V$ work by the hot post-shocked gas during the adiabatic Sedov–Taylor expansion phase of the SN blast \citep{Kim2015, Martizzi2015}, is instead injected into the neighboring gas cells \citep[see ][ for more details]{Marinacci2019}. Finally, since SN explosions are discrete events, the model mimics their discrete nature by imposing a time-step constraint for each stellar particle based on its age, such that the expectation value for the number of SN events per time-step is of order unity \citep{Hopkins2018, Marinacci2019}.

\subsubsection{Early stellar feedback}
\label{sec:esf}
Even though the feedback model introduced in \cite{Kannan2020b} and briefly outlined here has been successful in reproducing the correct star-formation rates  in isolated disk galaxy simulations \citep{Kannan2020b, Kannan2021}, going to a cosmological zoom-in setup results in larger than expected stellar mass of galaxies. This might either be due to some missing additional physics like cosmic rays \citep{CR2016}, magnetic fields \citep{Marinacci2016}, Lyman-alpha radiation pressure \citep{Smith2017,Kimm2018,Nebrin2024}, unresolved turbulence \citep{Semenov2016} or (despite our best efforts) an inability of the simulations to properly resolve/model feedback processes like photoheating \citep{Smith2022b}, radiation pressure from reprocessed IR radiation \citep{Murray2010} or even SN explosions \citep{1Katz1996}. To overcome this problem and match the stellar-mass-halo-mass estimates from abundance matching, we introduce an additional feedback channel labeled `Early Stellar Feedback' \citep[ESF; ][]{Stinson2013, Kannan2014b}. This feedback channel is motivated by recent observations that show that star-forming molecular clouds are disrupted by early feedback over very short time scales \citep{Kruijssen2019}. Similar to other empirical models developed to simulate this process \citep{Jeffreson2021, Keller2022}, we inject momentum into neighboring gas cells of newly formed stars for a duration of $5~\mathrm{Myr}$ after the star particle is formed. The amount of momentum injected per unit mass of the star particle is a free parameter which is set to
\begin{equation}
    \dot{p}_\mathrm{ESF}/m_\star = 1000 ~ \mathrm{km~s}^{-1}~\mathrm{Myr}^{-1} \, ,
\end{equation} 
to match the stellar-mass-halo-mass relations \citep{Moster2018, Behroozi2019} at high redshifts ($z\gtrsim3$). We note that this is about one \citep{Marinacci2019} to three times \citep{Kim2015} more momentum from SN than for a standard stellar population. A more thorough discussion of the impact of this feedback channel is discussed  in Section~\ref{sec:IESF}.

\subsubsection{Feedback coupling}
\label{sec:feedbackcoupling}
Following \citet{Marinacci2019}, the energy and momentum from the various feedback mechanisms are injected into the nearby gas cells using weight functions. This is achieved by calculating an effective number of neighbors for a star particle
\begin{equation}
    N_\mathrm{ngb} = \frac{4\pi}{3} h^3 \sum_i W(|\textbf{r}_i - \textbf{r}_s|, h)~,
    \label{eq:couple}
\end{equation}
where `$W$' is the cubic spline kernel \citep{Monaghan2015}, $h$ is the coupling radius, and $\textbf{r}_i$ and $\textbf{r}_s$ are the position vectors of the i-th gas neighbor and of the star particle, respectively. This equation is iteratively solved for $h$, such that a predetermined number of neighbors $N_\mathrm{ngb}$ is found. In addition, the coupling radius is limited in order to account for the fact that the energy/momentum from SN is not expected to have a strong impact on ISM properties beyond the superbubble radius \citep{Hopkins2018}. This radius is a function of the energy of the supernova and general ISM conditions like density, metallicity, and even the morphology of the ISM. To improve convergence \citep{Marinacci2019} we use a constant value of the limiting radius of $2~\mathrm{kpc}$ for our fiducial model. This value was chosen so that enough of the feedback energy is coupled to regulate the star-formation rate while allowing realistic superbubble sizes to form in the simulations.

Once the coupling radius has been determined, weights are defined so that each neighboring gas cell receives energy and momenta proportionally to the fraction of the $4\pi$ solid angle that it covers as seen from the star.  \citet{Marinacci2019} employs two distinct coupling radii and two different corresponding weights for the mass and feedback injections, respectively. While the feedback weights are calculated by considering only the gas particles within the limiting radius, the injection of ejected mass is not limited in terms of the radius to which it can be coupled. This differential weighting can artificially increase the total injected energy and momentum, particularly when only a few cells are within the limiting radius. In such cases, these cells receive a significant amount of additional energy and momentum while acquiring very little extra mass. This results in a boosted velocity kick, thus artificially enhancing the feedback injected into these cells. Consequently, this can lead to unusually large superbubble sizes and gas velocities. To address this issue, the weighting scheme for both mass and feedback is computed using the same method. We calculate the weights for all cells within the coupling radius as outlined in the previous paragraph. However, we set the weights for feedback injection to zero for any cells beyond the limiting radius, effectively discarding that part of the feedback. This approach ensures that all mass is injected within the coupling radius, while the feedback energy and momentum are only injected into cells within the limiting radius. By using the same weighting scheme for both mass and feedback, we eliminate the artificially boosted velocity injections.

\subsection{Large-scale radiation fields}
\label{sec:uvb}
External radiation fields can reduce the fraction of baryons that dark matter halos can retain, suppressing galaxy growth, especially at the low-mass end \citep[see for e.g. ][]{Okamoto2008}. Previous works have shown that this effect can reduce the number densities of these small galaxies \citep{Wu2019a}, which in turn reduces the number of photons available for reionizing the Universe \citep{Finkelstein2019}. Therefore, modeling this complex cyclical feedback loop requires radiation hydrodynamics coupled to accurate models for galaxy formation \citep{Pawlik2017, Borrow2023}. 

\begin{figure*}
    \includegraphics[width=0.999\textwidth]{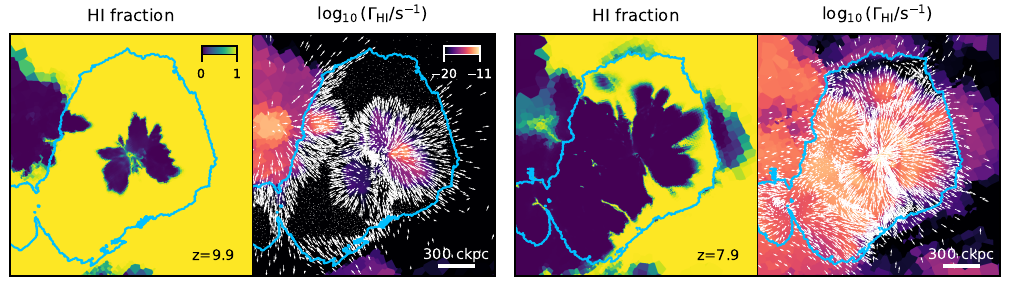}
    \caption{Treatment of the boundary conditions for the radiation field at the edge of the high-resolution region (marked by the blue contour) in the \mbox{$\mathrm{m}10.4\_4\mathrm{x}$} simulation at $z=9.9$ (left two panels) and $z=7.9$ (right two panels). The maps show the \ion{H}{I} fraction and the log of the \ion{H}{I} photoionization rate ($\Gamma_\ion{H}{I}$) as indicated. The white arrows in the \ion{H}{I} photoionization rate  plot represent the effective net velocity of the radiation field (photon flux divided by photon density). The plot clearly shows that the high-intensity external ionizing source present in the \thesanone simulation, shown on the left of the zoom-in region, reionizes part of the targeted high-resolution region, outside-in, at early times ($z=9.9$), before the local radiation field takes over and ionizes the majority of the region by $z=7.9$.}
    \label{fig:rtbound}
\end{figure*}

Despite this, most cosmological simulations typically use a time-varying but spatially uniform ultraviolet background \citep[UVB; ][]{FG09, HM2012, P19, CAFG2020} to simulate the impact of the large-scale radiation fields \citep[see for e.g. ][]{Vogelsberger2014, Schaye2015, Hopkins2018}. Typically, these models have a sharp transition between a fully neutral and a fully ionized Universe, which is necessitated by the spatial uniformity assumed in these models. They, therefore, miss the patchiness of reionization and corresponding variation in the radiation intensity \citep[although see][for recent efforts to include non-uniform UVBs]{Bird2022}. Moreover, the coupling between the radiation field and gas is modeled using a simple approximation that assumes that high-density gas cells/particles are self-shielded against external radiation \citep{Rahmati2013}. While this approximation is valid at low redshifts, where all gas will have been exposed to such radiation fields for billions of years, the local large-scale radiation environment is essential to model galaxies at $z>5$. Therefore, simulations aiming to model the formation and evolution of primordial galaxies must strive to include a realistic coupling between galaxies and a patchy, time-varying radiation field intensity. This is especially crucial in the current \jwst era, as we are beginning to observe the properties of high-redshift galaxies that are forming during reionization, including specific observables that are sensitive to IGM scales such as \lya emission and the relation between the sizes of ionized bubbles and the properties of galaxies within them \citep{Saxena2023}.

Large-volume, galaxy formation simulations with coupled radiation hydrodynamics currently provide the best description of the patchy and time-varying radiation field \citep{Gnedin2014, Rosdahl2018, Ocvirk2020, Kannan2022}. Unfortunately, zoom-in simulations (like the ones presented in this work) focus on simulating the properties of a single selected galaxy/dark matter halo and its immediate environment. They do not have the ability to model the astrophysics of large-scale structures, and therefore, do not contain any information about external radiation fields. Fortunately, the galaxies in the \thzoom project are selected from the parent \thesan simulations.  So, the large-scale structure, the properties of the galaxies, and the evolution of the radiation field around the selected halos have already been modeled. We can, therefore, follow the propagation of radiation across the boundary of the high-resolution region to ensure incoming radiation flows present in the parent \thesan simulation are seen by the high-resolution region of the zoom-in run. This allows for a more realistic treatment of reionization for objects that are inefficient sources of ionizing radiation and reionize outside-in. We use maps of the radiation field of the parent simulation that were saved with high cadence, that are interpolated in space and time to set the radiation field outside of the high-resolution region at each timestep. Inflowing radiation is then propagated into the high-resolution region using \areport. At low redshifts, after the final output of the parent simulation ($z=5.5$), we smoothly switch to setting the external radiation field based on a homogeneous UVB model from \citet{FG09}.

Fig.~\ref{fig:rtbound} illustrates this scheme for  the $\mathrm{m}10.4\_4\mathrm{x}$ simulation (see section~\ref{sec:init} for more details). The  two color-maps show the \ion{H}{I} fraction and the log of the \ion{H}{I} photoionization rate ($\Gamma_\ion{H}{I}$) at $z=9.9$ (left two panels) and $z=7.9$ (right two panels), as indicated. The blue contour marks the edge of the high-resolution region, while the white arrows represent the effective net velocity of the radiation field (photon flux divided by photon density for the $13.6-24.6$ eV bin). The high-intensity external ionizing source present in the \thesanone simulation, shown on the left of the zoom-in region, reionizes part of the targeted high-resolution region, outside-in, at early times ($z=9.9$). 

However, the local radiation field takes over and ionizes the majority of the region by $z=7.9$. Due to differences in galaxy formation physics between the original \thesan simulations and the current project, the re-simulated galaxy is a less efficient ionizing source than in the parent simulation. At $z=7.9$, it has produced a smaller ionized bubble, visible to the right of the galaxy, while on the left, the bubble has already merged with the external radiation field. Additionally, the disconnected ionized region observed just beyond the high-resolution zoom-in area in the top right of the plot is a consequence of the radiation field extending further out in the parent \thesan simulation. This radiation does not flow back into the zoom-in simulation, so the self-consistent expansion of the ionized bubble from the galaxy outwards remains unaffected in the high-resolution area. While this difference means that the incoming radiation field will vary slightly from what it would be if the entire box were simulated using the \thzoom model, it is currently the best approach available to model the patchy reionization process in zoom-in setup.

\subsection{Initial conditions}
\label{sec:init}

The dark matter halos of zoomed-in galaxies have been drawn from the parent dark matter only counterpart of the fiducial simulation in the \thesan project \citep{KannanThesan, Garaldi2022, Smith2022, ThesanDR} called \thesandarkone. It has a comoving boxsize of $95.5~\mathrm{cMpc}$ and a dark matter particle mass resolution of $3.7 \times 10^6~\Msun$. The cosmological parameters from, \citet{Planck2015_cosmo} are used (more precisely, the one obtained from their \texttt{TT,TE,EE+lowP+lensing+BAO+JLA+H$_0$} dataset), \ie $H_0 = 100h$ with $h=0.6774$, $\Omega_\mathrm{m} = 0.3089$, $\Omega_\Lambda = 0.6911$, $\Omega_\mathrm{b} = 0.0486$, $\sigma_8 = 0.8159$, and $n_s = 0.9667$, where all the symbols have the usual meaning. Objects with a wide range of masses contribute to cosmic reionization and exquisite resolution is required to accurately follow the internal structure of galaxies and the escape of ionizing photons. To best meet both requirements, we simulate galaxies over a wide mass range at three different resolution levels $4$x, $8$x, and $16$x, with linear spatial resolution improved by these factors, corresponding to $4^3$, $8^3$, and $16^3$ better mass resolution than the large volume \thesanone simulation. Table~\ref{table:res} lists the effective (total volume equivalent) particle number (second column), the dark matter (third column) and gas (fourth column) particle mass and the gravitational softening length for dark matter and stars (fifth column), and the minimum value of the adaptive softening length of the gas particles (sixth column) for the three different resolution levels used in this work. 

 \begin{table}
	\centering
	\caption{\textup{\thzoom simulation suite: From left to right, the columns indicate the name of the resolution level, effective (total volume-equivalent) number of particles, the mass of the high-resolution dark matter and gas particles, the softening length of star and dark matter particles and the minimum softening length for gas cells.}}
	\label{table:res}
 \small\addtolength{\tabcolsep}{-1.0pt}
	\begin{tabular}{llcccccccc} 
		\hline
        \vspace{-0.25cm}\\
		Name & $N_\mathrm{particles}^\mathrm{eff}$ & $m_\mathrm{DM}$ & $m_\mathrm{gas}$ & $\epsilon_\mathrm{DM, stars}$ & $\epsilon_\mathrm{gas}^\mathrm{min}$\vspace{0.05cm}\\  
		& & [$\Msun$] & [$\Msun$] & [cpc] & [cpc]\vspace{0.05cm}\\
		\hline
        \vspace{-0.2cm}\\
		16x & $2 \times 33600^3$  & $7.62 \times 10^2$ & $ 1.42 \times 10^2$ & $138.39$ & $17.30$\vspace{0.05cm}\\
		8x & $2 \times 16800^3$ & $6.09 \times 10^3$ & $1.14 \times 10^3$ & $276.79$ & $34.60$ &\vspace{0.05cm}\\
		4x & $2 \times 8400^3$ & $4.87 \times 10^4$ & $9.09 \times 10^3$ & $553.59$ & $69.20$\vspace{0.05cm}\\
		\hline
	\end{tabular}
\end{table}

\begin{table*}
	\centering
	\caption{\textup{The complete set of \thzoom simulations: The first column shows the names of the simulated galaxies (with the number referring to the logarithmic mass of the parent simulation object at $z=3$ to give a rough indication of the size of the object), while the second column is the number of the corresponding FOF group in the halo catalog of \thesandarkone at $z=3$ that is being  resimulated. The mass of the group (within $R_{200c}$) in units of solar masses is shown in the third column. The remaining columns indicate the resolution level and the numerical model utilized for the simulation. A check mark ($\checkmark$) denotes completion of the simulation, while an ‘$\text{\sffamily x}$’ indicates that the halo was not simulated at that specific resolution level and with that numerical model.}}
	\label{table:simtab}

	\begin{tabular}{lcccccccccccccc} 
		\hline
        \vspace{-0.22cm}\\
  \label{tab:sims}
 
  Simulation&Group Number&M$_\mathrm{group}$ at $z=3$&4x&8x&16x&4x & 4x & 4x & 8x & 8x & 8x & 8x & 8x \vspace{0.05cm}\\
  
  & \thesandarkone & [$\Msun$] &  &  &    &noESF & noRlim & varEff & noESF & noRlim & varEff & UVB & lateUVB \vspace{0.05cm}\\
 
  \hline
  \vspace{-0.2cm}\\
  m$13.0$ & 2 &  $8.93 \times 10^{12}$ & \checkmark & \text{\sffamily x} & \text{\sffamily x} & \text{\sffamily x} & \text{\sffamily x} & \text{\sffamily x} & \text{\sffamily x} & \text{\sffamily x} & \text{\sffamily x} & \text{\sffamily x} & \text{\sffamily x}\vspace{0.05cm}\\
    m$12.6$ & 39 &  $4.07 \times 10^{12}$ & \checkmark & \text{\sffamily x} & \text{\sffamily x} & \text{\sffamily x} & \text{\sffamily x} & \text{\sffamily x} & \text{\sffamily x} & \text{\sffamily x} & \text{\sffamily x} & \text{\sffamily x} & \text{\sffamily x}\vspace{0.05cm}\\
      m$12.2$ & 205 & $1.58 \times 10^{12}$ & \checkmark & \text{\sffamily x} & \text{\sffamily x} & \text{\sffamily x} & \text{\sffamily x} & \text{\sffamily x} & \text{\sffamily x} & \text{\sffamily x} & \text{\sffamily x} & \text{\sffamily x} & \text{\sffamily x}\vspace{0.05cm}\\
  m$11.9$ & 578 & $7.70 \times 10^{11}$ & \checkmark & \text{\sffamily x} & \text{\sffamily x} & \checkmark & \text{\sffamily x} & \text{\sffamily x} &  \text{\sffamily x}& \text{\sffamily x} & \text{\sffamily x} & \text{\sffamily x} & \text{\sffamily x}\vspace{0.05cm}\\
  m$11.5$ & 1163 &  $3.28 \times 10^{11}$ & \checkmark & $\text{\sffamily x}$ & \text{\sffamily x} & \checkmark & \text{\sffamily x} & \text{\sffamily x} &  \text{\sffamily x}& \text{\sffamily x} & \text{\sffamily x} & \text{\sffamily x} & \text{\sffamily x}\vspace{0.05cm}\\
  m$11.1$ & 5760 & $1.40 \times 10^{11}$ & \checkmark & $\checkmark$ & \text{\sffamily x} & \checkmark & \checkmark & \checkmark & \text{\sffamily x}& \text{\sffamily x} & \text{\sffamily x} & \text{\sffamily x} & \text{\sffamily x}\vspace{0.05cm}\\
  m$10.8$ & 10304 & $5.93 \times 10^{10}$ & \checkmark & \checkmark & \text{\sffamily x} & \checkmark & \text{\sffamily x}& \text{\sffamily x} & \text{\sffamily x}& \text{\sffamily x} & \text{\sffamily x} & \text{\sffamily x} & \text{\sffamily x}\vspace{0.05cm}\\ 
  m$10.4$ & 33206 & $2.53 \times 10^{10}$ & \checkmark & \checkmark & \text{\sffamily x} & \checkmark & \checkmark & \checkmark &  \checkmark & \checkmark & \checkmark  & \checkmark & \checkmark\vspace{0.05cm}\\ 
  m$10.0$ & 37591 & $1.07 \times 10^{10}$ & \checkmark & \checkmark &  \text{\sffamily x} & \text{\sffamily x} & \text{\sffamily x} & \text{\sffamily x} & \text{\sffamily x} & \text{\sffamily x}& \text{\sffamily x} & \checkmark& \checkmark\vspace{0.05cm}\\ 
  m$9.7$ & 137030 & $4.58 \times 10^{9}$ & \checkmark & \checkmark & \checkmark & \checkmark & \checkmark & \checkmark  &  \checkmark & \checkmark & \checkmark & \checkmark& \checkmark\vspace{0.05cm}\\ 
  m$9.3$ & 500531 & $1.95 \times 10^{9}$ & \checkmark & \checkmark &  \checkmark & \text{\sffamily x} & \text{\sffamily x} & \text{\sffamily x} & \text{\sffamily x}& \text{\sffamily x} & \text{\sffamily x} & \checkmark& \checkmark\vspace{0.05cm}\\ 
  m$8.9$ & 519761 & $8.29 \times 10^{8}$ & \checkmark & \checkmark &  \checkmark & \text{\sffamily x} & \text{\sffamily x} & \text{\sffamily x} & \text{\sffamily x}& \text{\sffamily x} & \text{\sffamily x} & \checkmark& \checkmark\vspace{0.05cm}\\ 
  m$8.5$ & 2274036 & $3.51 \times 10^{8}$ & \checkmark & \checkmark &  \checkmark & \text{\sffamily x} & \text{\sffamily x} & \text{\sffamily x} & \text{\sffamily x}& \text{\sffamily x} & \text{\sffamily x} & \checkmark & \checkmark\vspace{0.05cm}\\
  m$8.2$ & 5229300 & $1.52 \times 10^{8}$ & \checkmark & \checkmark & \checkmark & \text{\sffamily x} & \text{\sffamily x} & \text{\sffamily x} &  \text{\sffamily x}& \text{\sffamily x} & \text{\sffamily x} & \checkmark& \
 \checkmark \vspace{0.05cm}\\
\hline
\vspace{-0.2cm}\\
Total & & 60 & 14 & 9 & 5 & 6 & 3 & 3 & 2 & 2 & 2 & 7 & 7 \vspace{0.05cm}\\
\hline
 \end{tabular}%
\end{table*}

The dark matter halos selected for resimulation from the \thesandarkone volume are identified at $z=3$ via the friends-of-friends  (FOF) algorithm \citep{Davis1985} using a linking length of 0.2 times the initial mean inter-particle distance. We select $14$ galaxies with halo masses ranging from $\sim10^8 - 10^{13}~\Msun$ at $z=3.$  They are selected so that we obtain reasonable
statistics with at least a few objects per dex in halo mass. The initial conditions of the selected halos are constructed using a  new zoomed-initial conditions code (Puchwein et al. in prep) which makes our runs numerically more efficient by optimizing the high-resolution region and the distribution of boundary
particles. The resolution levels of the latter can now have arbitrary shapes and are chosen based on the distance to the nearest edge of the high-resolution region. This
allows us to more efficiently concentrate the numerical effort on the region that is relevant for the scientific analysis. We have also increased the accuracy of the displacement calculation by making the matching in $k$-space between low- and high-resolution power more accurate and
using the appropriate Nyquist frequency as $k$-cutoff separately for each resolution level of the boundary particles (Puchwein et al. in prep). To reduce contamination by low-resolution particles, the selected high-resolution region at $z=3$ extends out to $4r_\mathrm{vir}$, where $r_\mathrm{vir}$ is the virial radius of the halo. All the simulations presented in this work have zero contamination within the virial radius of the target halo (in fact, for most simulations, there is no contamination within $2r_\mathrm{vir}$), and the mass distribution in the high-resolution region is in excellent agreement with the corresponding region of the parent \thesandarkone simulation, except for deviations on small scales that arise from the inclusion of baryonic physics.

\begin{figure*}
	\includegraphics[width=0.99\textwidth]{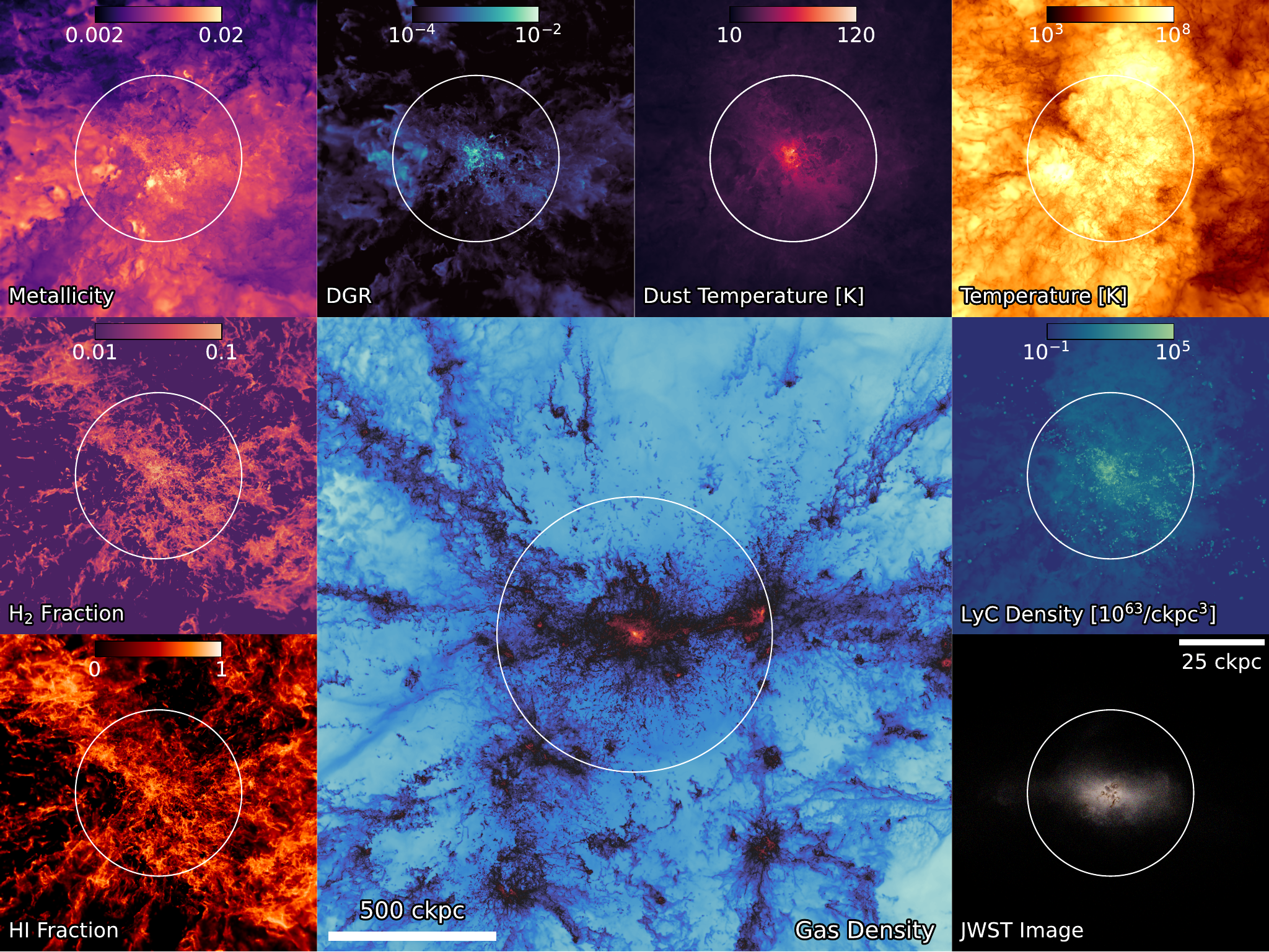}
    \caption{A qualitative illustration of the various physical quantities predicted by the \thzoom simulations at $z=3.2$. The central figure shows the large-scale ($2~\mathrm{cMpc}$) gas density distribution around the \target galaxy in the \mbox{m$12.6\_4\mathrm{x}$} simulation, with the white circle denoting its virial radius. The smaller figures showcase a smaller region of $100~\mathrm{ckpc}$ around the \target galaxy, and the circles indicate the half-mass radius of stars. Clockwise from bottom-left, we plot the \ion{H}{I} fraction, \ion{H}{$_2$} fraction, gas phase metallicity, dust-to-gas-ratio (DGR), dust temperature distribution, gas temperature distribution, the LyC photon density (combining all three LyC bands) and a mock \jwst image generated from the F277W, F356W, F444W bands. Except for the last image, all plots show the mass-weighted averages along the line of sight, which extends over the same length as the other two dimensions of the figure.}
    \label{fig:all}
\end{figure*}

\subsection{Simulations}

The complete set of \thzoom simulations is listed in Table~\ref{table:simtab}. The first column lists the names of the simulated galaxies, while the second column is the number of the
corresponding FOF group in the halo catalog of \thesandarkone at $z=3$.  The mass of the group in units of solar masses is outlined in the third column. The remaining columns indicate the resolution level and the numerical model utilized for the simulation. A check mark ($\checkmark$) denotes completion of the simulation, while an ‘$\text{\sffamily x}$’ indicates that the halo was not simulated at that specific resolution level and with that numerical model. The simulations were completed  down to $z=3$ so that we cover both the epoch of reionization as well as the redshift range in which the MUSE instrument provides detailed observations of Lyman-$\alpha$ emitters\footnote{The G2 halo was only run down to $z=6$ because our simulations lack blackhole feedback which has been shown to be extremely important at $z<6$ for these massive halos}. Due to the computational cost, simulating the most massive halos is only possible at lower resolution levels.  So we simulate all zoom-in regions, including the most massive halos at standard resolution (4x), halos up to $\mathrm{M}_\mathrm{halo} \sim 10^{11} ~ \Msun$ at increased resolution (8x) as well and only $\mathrm{M}_\mathrm{halo} \lesssim 10^{10} ~ \Msun$ at the highest-resolution (16x) level.

In addition to the fiducial simulations, the \thzoom suite also features additional runs designed to investigate the effect of varying numerical parameters and physical models on galaxy properties. They are:
\begin{itemize}
    \item \textbf{noESF}: The early stellar feedback channel described in Section~\ref{sec:esf} is not included.
    \item \textbf{noRLim}: The extent to which the feedback energy and momentum from young star particles are injected into the surrounding gas is effectively unrestricted. This is achieved by switching off the limiting radius for feedback coupling (see Section~\ref{sec:feedbackcoupling}) . 
    \item \textbf{varEff}: The star formation efficiency per free fall time ($\epsilon_\mathrm{ff}$, see Section~\ref{sec:sf} for more details) is changed from $100\%$ ($\epsilon_\mathrm{ff}=1$) to a variable value.  It starts off small with $\epsilon_\mathrm{ff} = 0.01$ at the threshold density of $n_\mathrm{H} = 10~\mathrm{cm}^{-3}$ and scales linearly with the density of the gas to reach a maximum value of $\epsilon_\mathrm{ff} = 1$ at $n_\mathrm{H} \geq 10^3~\mathrm{cm}^{-3}$. 
    \item \textbf{UVB}: The large-scale radiation field derived from the \thesan simulations is replaced by a spatially constant, time-varying UVB taken from \citet{FG09}. The gas cells feel this UV background in addition to radiation from local sources that are self-consistently modeled by the RHD solver \citep{Kannan2020b}. Like other cosmological simulations, the radiation intensity of the UVB is attenuated for the high-density gas using the self-shielding prescriptions outlined in \citet{Rahmati2013}.
    \item \textbf{lateUVB}: The large-scale field derived from the \thesan simulations is turned off until $z=5.5$, beyond which we switch to setting the external radiation field based on the homogeneous UVB model from \citet{FG09}. Local sources are modeled self-consistently.
\end{itemize}

In total, this gives
us a sample of 60 simulations, which are used to investigate the formation and evolution of high-redshift galaxies.

\begin{figure*}
	\includegraphics[width=0.99\textwidth]{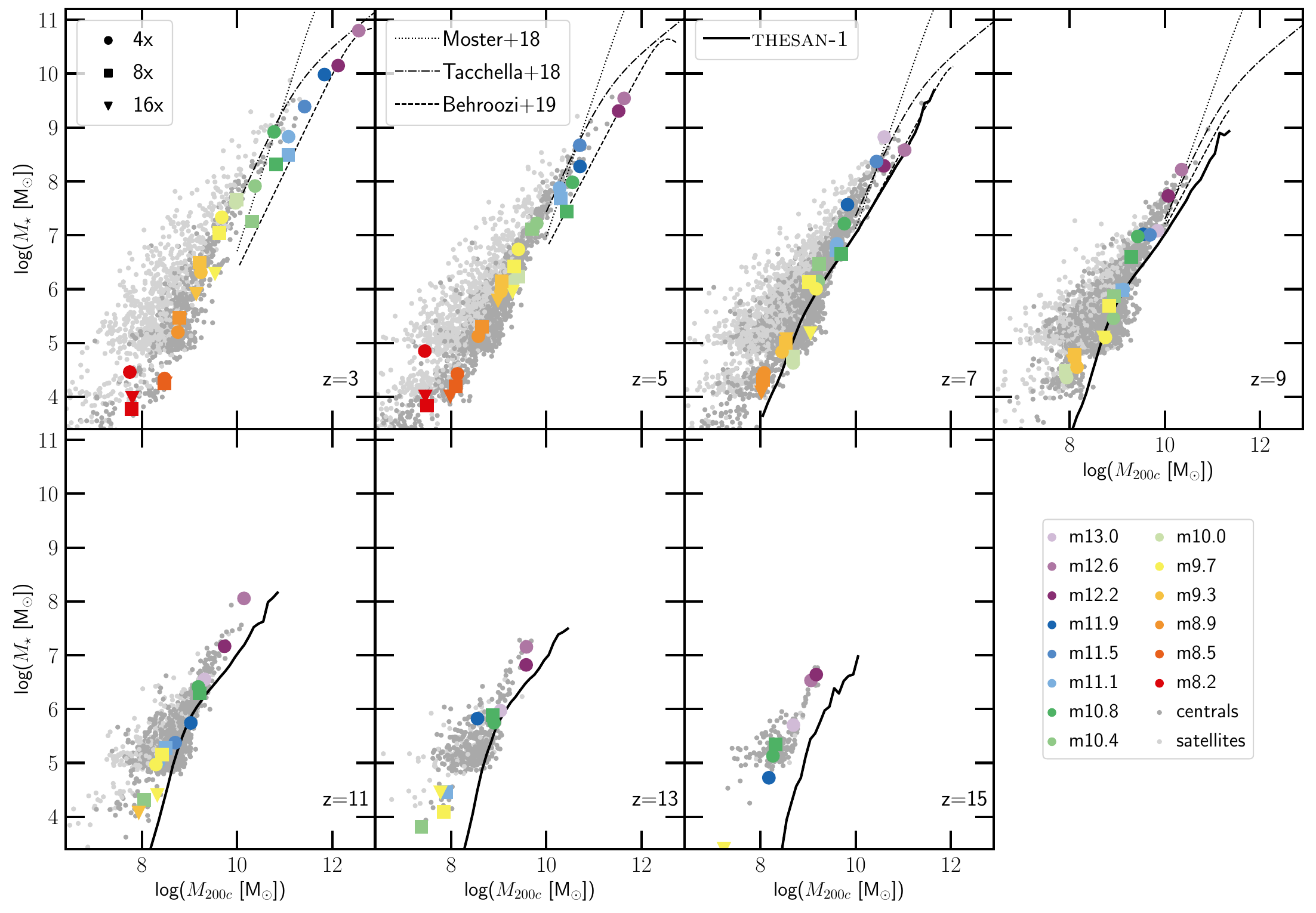}
    \caption{The stellar mass (within twice the stellar half mass radius)--halo mass (within $R_{200c}$) relation for the galaxies in the \thzoom simulations. The colored symbols show this relation for the \target galaxies, while the different marker styles indicate different resolution levels: $4$x, $8$x, and $16$x, indicated by circles, squares, and inverted triangles, respectively. The dark gray points are the \centrals, and the lighter gray points show this relation for the \satellite population. For comparison, we also show estimates from abundance matching techniques like UNIVERSEMACHINE \citep[][dashed curve]{Behroozi2019} and EMERGE \citep[][dotted curve]{Moster2018} and the empirical model by \citet[][dot-dashed curve]{Tacchella2018}. Finally, the solid black line plots the SHMR for the galaxies in the parent \thesanone \citep{KannanThesan} simulation. The stellar masses of \thzoom galaxies align with predictions from abundance matching results within the mass ($M_{200c} \gtrsim 10^{10}~\mathrm{M}_\odot$) and redshift ($z<10$) range they overlap.}
    \label{fig:smhm}
\end{figure*}

\subsection{Galaxy catalogs}

A total of 188 snapshots were written, every $10~\mathrm{Myr}$, from $z=16$ 
down to $z=3$. The dark matter halos are
identified using the friends-of-friends (FOF) algorithm using a linking length of 0.2 times the initial mean inter-particle distance. Stellar particles and gas cells are attached to these FOF primaries in a secondary linking stage. The SUBFIND-HBT algorithm \citep{Gadget4} is then used to identify gravitationally bound structures. These FOF and SUBFIND catalogs
accompany each snapshot output and contain a wide array of information about the detected DM halos and their gas and stellar properties.

We identify the target DM halo of the zoom-in simulation by requiring that:
\begin{itemize}
\item The distance between the center of the DM halo in the zoom-in simulation and \thesandarkone be less than $100~\mathrm{ckpc}$ at $z=3$.
\item The total mass difference between the zoom-in DM halo and DM halo in \thesandarkone is less than $0.5$~dex.
\item The zoom-in halo does not contain any low-resolution DM particles within the FOF group.
\end{itemize}

These criteria clearly identify the target galaxy for almost all halos in our simulation suite. Some of the lowest-mass halos occasionally have multiple subhalos that meet these criteria. In such cases, we select the one closest to the center of the parent halo in \thesandarkone. This DM halo's most massive galaxy (sometimes called the central galaxy) is then designated as the \textit{target} galaxy for the simulation. We trace the progenitors of the target galaxy over time using the SUBFIND-HBT algorithm and assign the most massive progenitor as the target galaxy at that redshift. This method enables us to consistently assign and trace the target galaxy throughout all redshifts. Additionally, all dark matter (DM) halos that do not contain any low-resolution DM, gas or star particles within $r_{200,c}$ are designated as uncontaminated and included in our analysis. The central galaxies of the uncontaminated DM halos are categorized as \textit{centrals}, while all other galaxies belonging to the FOF group are categorized as \textit{satellites}. Therefore, all \target galaxies are classified as \centrals, but not all \centrals are \target galaxies.

\section{Results}
\label{sec:results}

We begin with a qualitative illustration of the various physical quantities predicted by the \thzoom simulations (Figure~\ref{fig:all}). The large central figure plots the gas density distribution in a region of size $2~\mathrm{cMpc}$ around the \target galaxy of the \mbox{m$12.6\_4\mathrm{x}$} simulation, with the white circle denoting its virial radius ($r_\mathrm{vir}\sim 470~\mathrm{ckpc}$). We can clearly see the large-scale cosmic web with multiple filamentary structures feeding the high-density central regions that host galaxies. The surrounding smaller plots zoom-in on the central $100~\mathrm{ckpc}$ region around the \target galaxy with the white circles showing the half-mass radius of stars in the galaxy. Going clockwise from the bottom-left, we plot the neutral hydrogen fraction, the molecular hydrogen fraction, gas phase metallicity, the dust-to-gas ratio (DGR), dust temperature, gas temperature, the LyC photon density (combining all three EUV bands), and a mock \jwst image generated from the F277W, F356W, F444W bands by postprocessing the simulations using \textsc{skirt}.

\begin{figure*}
	\includegraphics[width=0.99\textwidth]{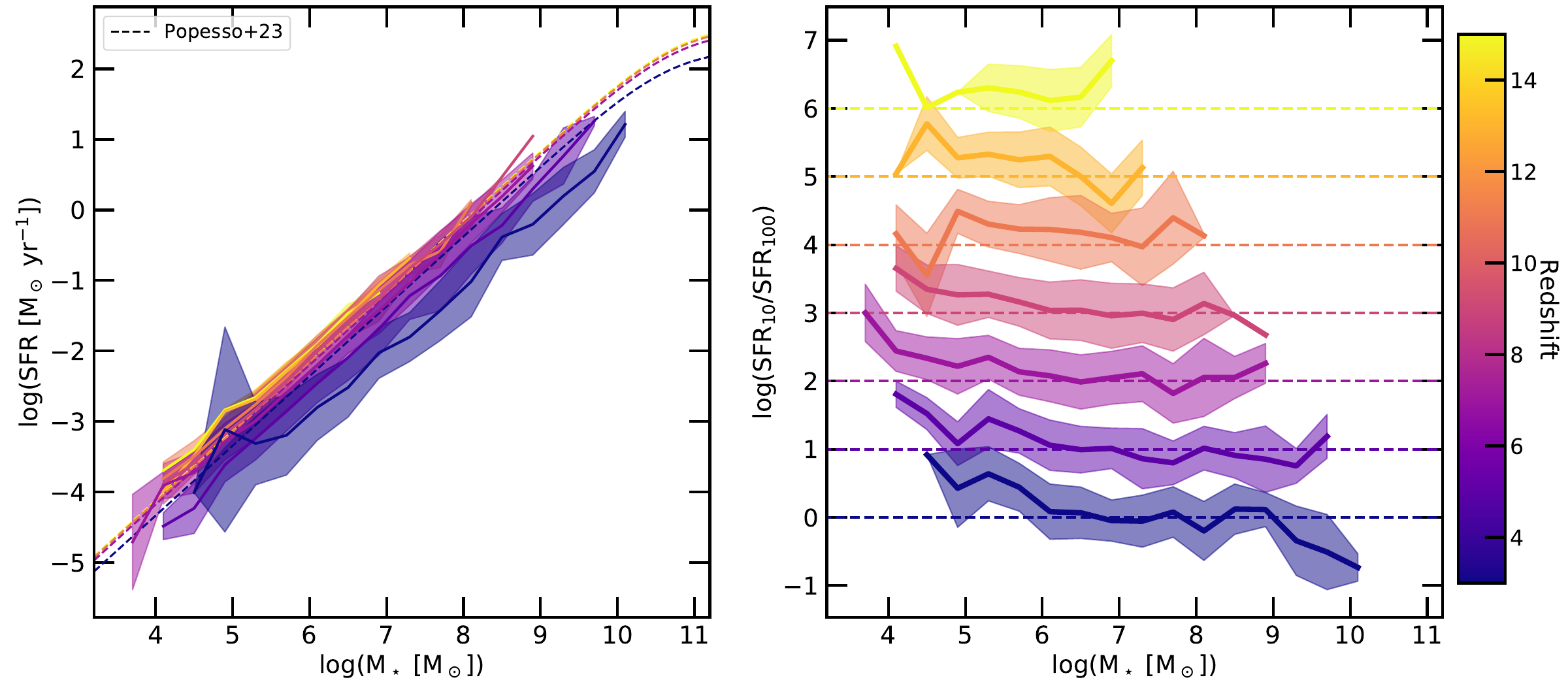}
    \caption{\textbf{Left Panel:} The star-formation rate (SFR) -- stellar mass relation for all the \central galaxies in the \thzoom simulations at $z=3-15$ in steps of $\Delta z=2$. The solid colored lines show the SFR averaged over $100~\mathrm{Myr} ~ (\mathrm{SFR_{100}})$ and the shaded region is the one sigma distribution around the mean. The simulated star-forming main sequence shows a much stronger evolution compared to the fits to the observed main sequence (colored dashed lines) outlined in \citet{Popesso2023}. \textbf{Right Panel:} The mean ratio of the SFRs averaged over $10~\mathrm{Myr}$ and $100~\mathrm{Myr}$ as a function of the stellar mass of the galaxy and redshift. The lines are shifted by $(z-3)/2$ for clarity. The horizontal dashed lines mark the zero point of the shifted values.
}
    \label{fig:sfr}
\end{figure*}

As expected, the neutral gas distribution is extended and more evenly distributed than the molecular gas in the galaxy.  The \HM distribution is clumpy, and the distribution peaks in the high-density regions with high dust content. This is because the typical \HM fractions are primarily set by competition between formation mediated by dust grains (sensitive to the local dust-to-gas ratio) and destruction by LW radiation, which is sensitive to the local radiation field and the gas density distribution. Similarly, the dust distribution is more clumpy and more concentrated in the higher-density regions, whereas the gas phase metals are more evenly spread out. The dust primarily forms in high-density gas because the deposition rate of gas-phase metals onto existing dust
grains is larger as the gas density increases (Eq.~\ref{eq:dgr}).  The DGRs also anti-correlate with the temperature of the gas because colder gas has a higher rate of dust formation, while hot gas ($T>10^4~\mathrm{K}$) can destroy dust by thermal sputtering. This anti-correlation is also seen in the \HI and \HM fractions because hot gas can lead to higher collisional ionization rates, leading to a destruction of neutral and molecular hydrogen phases. We note that the central region of the \target galaxy looks to have low metallicities, but highly dust enriched. The high dust enrichment and low metallicity indicates that a large fraction of the metal content has been deposited into dust. 

The dust temperature distribution depends on the composition and sizes of the dust grains, as well as the interstellar radiation field (ISRF). Most of the far-UV and visible light from stars is absorbed by dust and re-radiated in the IR regime. The map of ionizing emissivity is clustered and seems to correlate well with the prevalence of molecular gas. These sites are conducive for star-formation and emit copious amounts of LyC radiation during the first few million years of the stellar lifetime. The radiation is promptly absorbed by neutral gas close to the source due to their short mean free paths, giving the map its clumpy structure. Finally, the bottom right image shows a mock \jwst image generated from the F277W, F356W, F444W filters. The image was generated using the post processing Monte Carlo radiative transfer code \textsc{skirt} \citep{Camps2020} using the \textsc{toddlers} emission line library \citep{toddlers}. We refer the reader to the accompanying paper, Popovic et al., (in prep) for more details about the dust and emission line analysis of the \thzoom galaxies using \textsc{skirt}. We will note that the gas in these high-redshift galaxies seems to be a lot more clumpy and less structured than their low-redshift counterparts. This is an expected result of hierarchical structure formation, which predicts higher merger rates, larger gas fractions leading to Toomre-unstable disks \citep{Dekel2014, Gurvich2023, Hopkins2023} and turbulence due to feedback from high star-formation rates \citep{Li2025} at high-$z$. A more thorough description of the sizes and morphology of the simulated galaxies and their comparison to \jwst observations is made in an accompanying paper \citep{McClymont2025b}.

\subsection{Stellar properties}
We proceed by plotting the stellar-to-halo mass relation (SHMR), which quantifies the efficiency with which stars form in dark matter halos \citep{Moster2010}. Figure~\ref{fig:smhm} shows this relation for all the simulations run using the fiducial model by plotting the stellar mass of galaxies (within twice the stellar half mass radius) as a function of the total halo mass within $R_{200c}$, the radius which encloses matter with a mean density $200$ times the critical
density of the Universe. The different colored symbols plot this relation for the \target galaxies as indicated in the legend. The dark gray circles are the \centrals, while the lighter gray circles are the \satellites. For comparison, we also plot estimates from abundance matching techniques like UNIVERSEMACHINE \citep[dashed curve; ][]{Behroozi2019} and EMERGE \citep[dotted curve; ][]{Moster2018} and a simple empirical model that assigns a star-formation rate (SFR) to each dark matter halo based on the growth rate of the halo and a redshift-independent star formation efficiency \citep[dot-dashed curve; ][]{Tacchella2018}. Finally, the solid black curve shows the average stellar-to-halo-mass relation for the parent \thesanone simulation. We do not show the \thesan SHMR below $z<5$ because the simulations were only run down to $z=5.5$. 

At $z\lesssim9$, the stellar masses of the simulated target galaxies with $M_\mathrm{halo} \gtrsim 10^{10}~\Msun$ generally lie in between the EMERGE \citep{Moster2018} and UNIVERSEMACHINE \citep{Behroozi2019} abundance matching estimates. They are also in good agreement with the values from the semi-empirical model of \citet{Tacchella2018}. Due to the challenges associated with measuring the stellar content of low-mass galaxies, particularly at high redshifts, the abundance matching results do not extend below $M_\mathrm{halo} \lesssim 10^{10}~\Msun$ or above $z=10$. Consequently, it is difficult to validate the predictions at the low-mass end. However, we note that the lower mass galaxies in the \thzoom simulation set appear to be in good agreement with the parent \thesanone simulation, at $z=7-11$, but they exhibit slightly higher stellar masses at $z\geq13$. 

The \centrals occupy a similar parameter space as the \target galaxies, but the satellite population has a generally higher stellar mass for a given halo mass, probably due to the fact that it is not possible to effectively define their virial radius once they fall into the host dark matter halo. The three different (\fx, \ex, and \sx) resolution simulations predict similar stellar masses for most halos in the \thzoom suite. There are, however, a few exceptions like \mbox{m$10.4$}, \mbox{m$9.7$} and \mbox{m$8.2$} that show a difference of about $\lesssim 0.3~\mathrm{dex}$ in stellar mass of \target galaxies between the various resolution levels. There is also a trend of higher-resolution simulations forming fewer stars in general. Overall, the SMHM relation shows that the fiducial model is able to reproduce the observed star formation efficiency over a wide halo mass and redshift range. We note that this is only possible with the inclusion of ESF, which makes feedback more efficient and helps reduce the stellar masses of the higher-mass galaxies in our simulation suite (see Sec.~\ref{sec:IESF} for more details).

We now turn our attention to the simulated star-forming main sequence shown in the left panel of Figure~\ref{fig:sfr}. It plots the star-formation rate of all the \centrals as a function of their stellar mass. The colors indicate different redshifts from $z=3-15$, with lighter colors depicting lower redshifts as shown by the colorbar. The solid colored lines show the SFR averaged over $100~\mathrm{Myr} ~ (\mathrm{SFR_{100}})$ and the shaded region is the one sigma distribution around the mean. The dashed colored lines are the fits to the observed main sequence of star-forming galaxies from \citet{Popesso2023}. The redshift evolution of the simulated main sequence shows a much stronger evolution compared to the observationally derived fits.  This probably arises due to observational biases, where the slope is artificially flattened due to stellar mass incomplete samples. We refer the reader to the accompanying paper \citet{McClymont2025a} for a more thorough discussion on this topic.  The right panel of Figure~\ref{fig:sfr} plots the ratio $\mathrm{SFR}_{10}$/$\mathrm{SFR}_{100}$, which acts as an  indicator of the star formation history and stochasticity of the current star formation activity in the galaxy. The solid lines show the median relation while, the shaded regions are the one sigma distribution. To improve clarity, we offset the lines by a factor of $(z-3)/2$ and mark the zero point of this offset relation by horizontal dashed lines. At $z\geq10$, the median relation lies above the zero point, indicating that the most recent star formation is larger than the star-formation rate averaged over $100~\mathrm{Myr}$, which points to generally increasing star formation histories at these redshifts. At later times ($z\lesssim4$) the higher-mass galaxies (M$_\star \gtrsim 10^8$) tend to have generally declining SFRs, while the lower-mass galaxies are still rapidly increasing their stellar masses.

\begin{figure}
	\includegraphics[width=0.99\columnwidth]{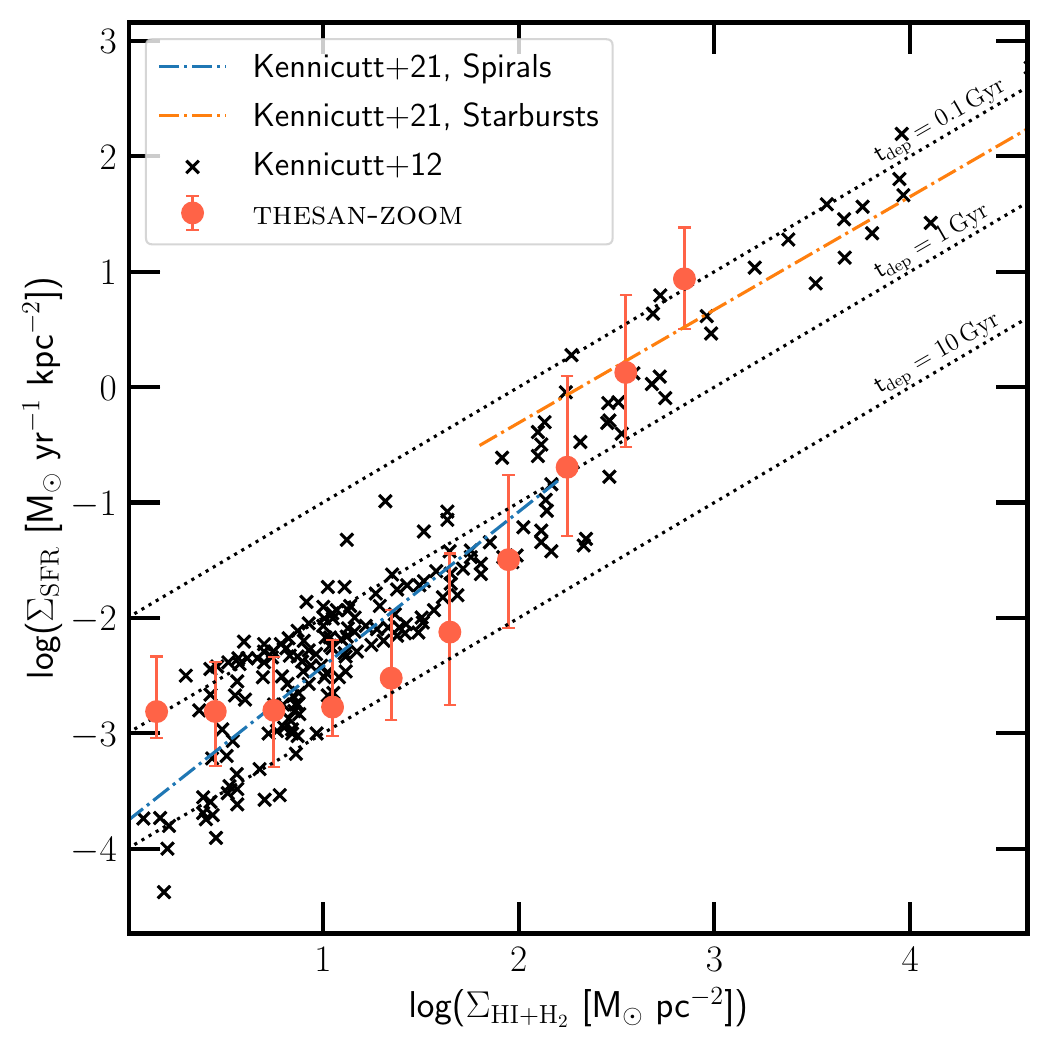}
    \caption{The star-formation rate surface density plotted as a function of the total neutral and molecular gas surface density for the \target galaxies, at $z=3$ (red points). The observational estimates for the Milky-Way and other nearby galaxies \citep{Kennicutt2012} are plotted as black crosses while the blue and orange lines indicate the relation derived from observations of local spirals and star-bursting galaxies \citep{Kennicutt2021}.  The gas depletion times are  $\tau_\mathrm{dep}\gtrsim 1~\mathrm{Gyr}$, in low gas surface density environments ($\Sigma_\mathrm{HI+H_2} \lesssim 300~\Msun~\mathrm{pc}^{-2}$), but decreases to about $0.1~\mathrm{Gyr}$ at higher surface densities. }
    \label{fig:KS}
\end{figure}

Next, we look at the relationship between the SFR of the galaxy and its cold gas content by plotting the star formation surface density as a function of the surface density of neutral and molecular gas (\ion{H}{I} + \ion{H}{$_2$}) at $z=3$, as shown in Figure~\ref{fig:KS}. The red circles are the simulation estimates calculated for the \target galaxies (with \fx resolution) in \thzoom by averaging over $1~\mathrm{kpc}^2$ patches. SFR here is averaged over the last $10$\,Myr. We combine measurements of patches from all galaxies and the errorbars show the one sigma variation (including both pixel-by-pixel and galaxy-by-galaxy variations) in the relation. The diagonal dotted lines are the locus of constant gas depletion times (defined as $\tau_\mathrm{dep} = \Sigma_\mathrm{SFR}/\Sigma_\mathrm{HI + H_2}$) with three different times of $0.1~\mathrm{Gyr}$, $1~\mathrm{Gyr}$ and $10~\mathrm{Gyr}$ shown as labeled. The black crosses are a compilation of observational results for the galaxies in the local Universe \citep{Kennicutt2012}. The blue dashed, and orange dashed lines are separate power law fits to the non-starbursting (or spiral) and starbursting galaxies from \citet{Kennicutt2021}. Interestingly, they show very different slopes, with the non-starbursting galaxies showing a steep slope of $n=1.34 \pm 0.07$ while the starbursting galaxies have a shallower slope of  $n=0.98 \pm 0.07$. However, there is a pronounced offset in the zero-point ($\sim0.6~\mathrm{dex}$) for starbursting galaxies to higher SFR surface densities.

The simulated gas depletion times frequently exceed $1~\mathrm{Gyr}$, particularly for gas surface densities below $100~\Msun~\mathrm{pc}^{-2}$. The flattening of the relationship at low surface density is an artifact arising from limited numerical resolution, which results in the presence of only a single young ($<10~\mathrm{Myr}$) stellar particle (with mass of $\sim 9.09 \times 10^3~\Msun$) per pixel. As the gas surface density increases, the depletion times decrease sharply, reaching values as low as $0.1~\mathrm{Gyr}$ for surface densities near $300~\Msun~\mathrm{pc}^{-2}$. This behavior closely aligns with the observed trend of reduced depletion times in starburst galaxies characterized by high gas surface densities. However, the simulations indicate a more gradual transition to lower depletion times in contrast to the abrupt changes observed empirically. This indicates that the differences in Schmidt laws seen between normal non-starbursting galaxies  and starbursts merely reflect the extremes of a more continuous underlying change in the relation rather than a true bimodality. Finally, we note that the simulated KS relation seems to be independent of redshift, numerical resolution, star-formation efficiency per free-fall time, or the value of the limiting radius \citep{Wang2025}. The only exception is that runs without ESF show very short gas depletion times (see Section~\ref{sec:IESF} for more details). This is an expected result if the turbulent momentum dissipation and energy injection from feedback are in equilibrium \citep{Thompson2005, Ostriker2011, FG13, Shen2025}.

\subsubsection{Impact of early stellar feedback}
\label{sec:IESF}
The previous plots show that the fiducial feedback model effectively reproduces the stellar masses and star-formation rates of galaxies across a broad range of halo masses and redshifts ($z=3-15$). However, as discussed in section~\ref{sec:esf}, an additional early stellar feedback channel is required to achieve this good match. This channel injects momentum into neighboring gas cells of newly formed
stars for a duration of $5$\,Myr immediately after the star particle is formed. The amount of momentum injected is tuned to match the high-redshift stellar-to-halo mass relation.

\begin{figure}
	\includegraphics[width=0.99\columnwidth]{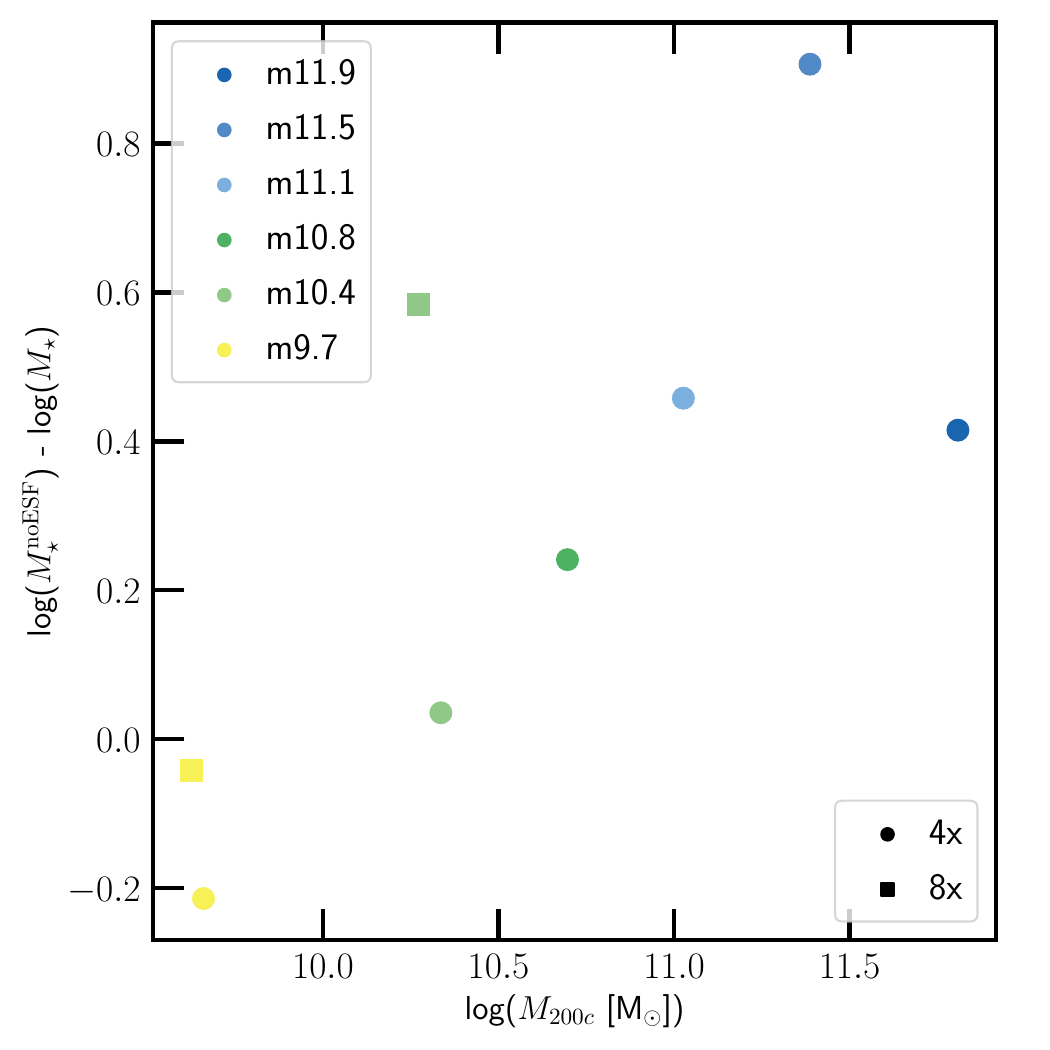}
    \caption{The difference in stellar masses between the \target galaxies of ``noESF'' and ``fiducial'' models at $z=3$. We show this result for all fiducial simulations that have a ``noESF'' counterpart (see Table~\ref{tab:sims} for more details). The results for the $4$x and $8$x resolution runs are indicated as circles and squares, respectively. Early stellar feedback is crucial for reducing the stellar masses of galaxies, especially at the high mass end.}
    \label{fig:esf}
\end{figure}

Figure~\ref{fig:esf} illustrates the effect of this additional feedback by plotting the difference in stellar mass in simulations with (fiducial) and without (noESF) early stellar feedback as a function of halo mass at z=3. The plot reveals a clear correlation between the efficacy of early stellar feedback and the halo mass of galaxies. Notably, ESF is significantly more effective at reducing the stellar content of higher-mass galaxies.  At \fx resolution, ESF can decrease the stellar mass by more than $0.4$~dex (approximately a factor of 2.5) in galaxies with $M_\mathrm{halo} \geq 10^{11}~\Msun$. In contrast, its impact on lower-mass galaxies is less pronounced, with reductions of $\leq 0.2$~dex. In fact, for the low-mass m$9.7$ galaxy, the stellar mass is higher in the fiducial simulation compared to the ``noESF'' run. The \ex resolution runs show a similar behavior of ESF being more effective at reducing the stellar masses of high-mass galaxies. Interestingly, for a given halo mass, ESF is able to regulate star formation more efficiently in the \ex runs compared to the \fx simulations. For example, the difference in stellar mass between m$10.4\_4\mathrm{x}$ and m$10.4\_4\mathrm{x}\_\mathrm{noESF}$ runs is minimal, while it is about $0.6$~dex for the m$10.4\_8\mathrm{x}$ and m$10.4\_8\mathrm{x}\_\mathrm{noESF}$ runs. The reason for this mass and resolution dependence is beyond the scope of this work and will be investigated more thoroughly in a follow-up paper.

Without ESF, the feedback model is unable to effectively disperse high-density star-forming gas clouds. This model was previously only tested in isolated simulations of galaxies with properties similar to the Milky Way and the Large Magellanic Cloud \citep{Kannan2020b, Kannan2021}. These simulations had low gas fractions ($\leq 0.2$) and, averaged over kpc scales, only gas surface densities less than $\sim 30~\Msun~\mathrm{kpc}^{-2}$ were probed \citep{Marinacci2019}. However, the conditions in high-redshift galaxies are known to be quite different. The gas fractions and surface densities are generally higher, and the metallicities are lower \citep{Car2013}. In these conditions, the ability of stellar feedback to self-regulate decreases. This is partly due to the slight decrease in the injected terminal momentum from a SN remnant, which scales very weakly with density as $\sim \rho^{-0.17}$ \citep{Kim2015, Martizzi2015}. This weak scaling arises because the increase in cooling is largely offset by the SN remnant sweeping up more ambient ISM mass at higher densities. More importantly, at sufficiently high surface densities ($\Sigma\gtrsim 1000 \, \mathrm{M}_\odot \, \mathrm{pc}^{-2}$), various standard stellar feedback mechanisms, collectively, do not output enough momentum to halt the  gravitational collapse of the gas cloud, leading to the cloud-scale SFE reaching unity \citep{Grudic2018, Grudic2019, Menon2024}.

Early stellar feedback mechanisms like photoheating and stellar winds also become less effective in these high-density, low-metallicity environments. The impact of photoheating reduces because it can only induce velocities of order of the sound speed in the ionized gas ($\sim 10~\mathrm{km}~\mathrm{s}^{-1}$) \citep{Rosdahl2015, Kannan2020a} which is lower than the velocities required to disrupt high-density ($\geq 10^3~\mathrm{cm}^{-3}$) giant molecular clouds \citep{Kim2021}. Radiation pressure, both from direct  UV and reprocessed IR radiation, can potentially disrupt these massive clouds, but recent simulations show that high-density GMCs only become super-Eddington once high star formation efficiencies of $\sim 80\%$ are reached \citep{Menon2023}. Similarly, the mechanical energy output of stellar winds reduces as the metallicity of the star decreases. These winds are driven by radiation pressure on the resonant absorption lines of metals found in the atmosphere of stars like carbon and nitrogen \citep{Castor1975}. Stellar evolutionary models predict that the average mechanical output in the form of stellar winds is about two orders of magnitude lower for a young stellar population with a metallicity of $\sim 0.02~\Zsun$ compared to solar metallicity stars \citep{SB99}.

\begin{figure}
	\includegraphics[width=0.99\columnwidth]{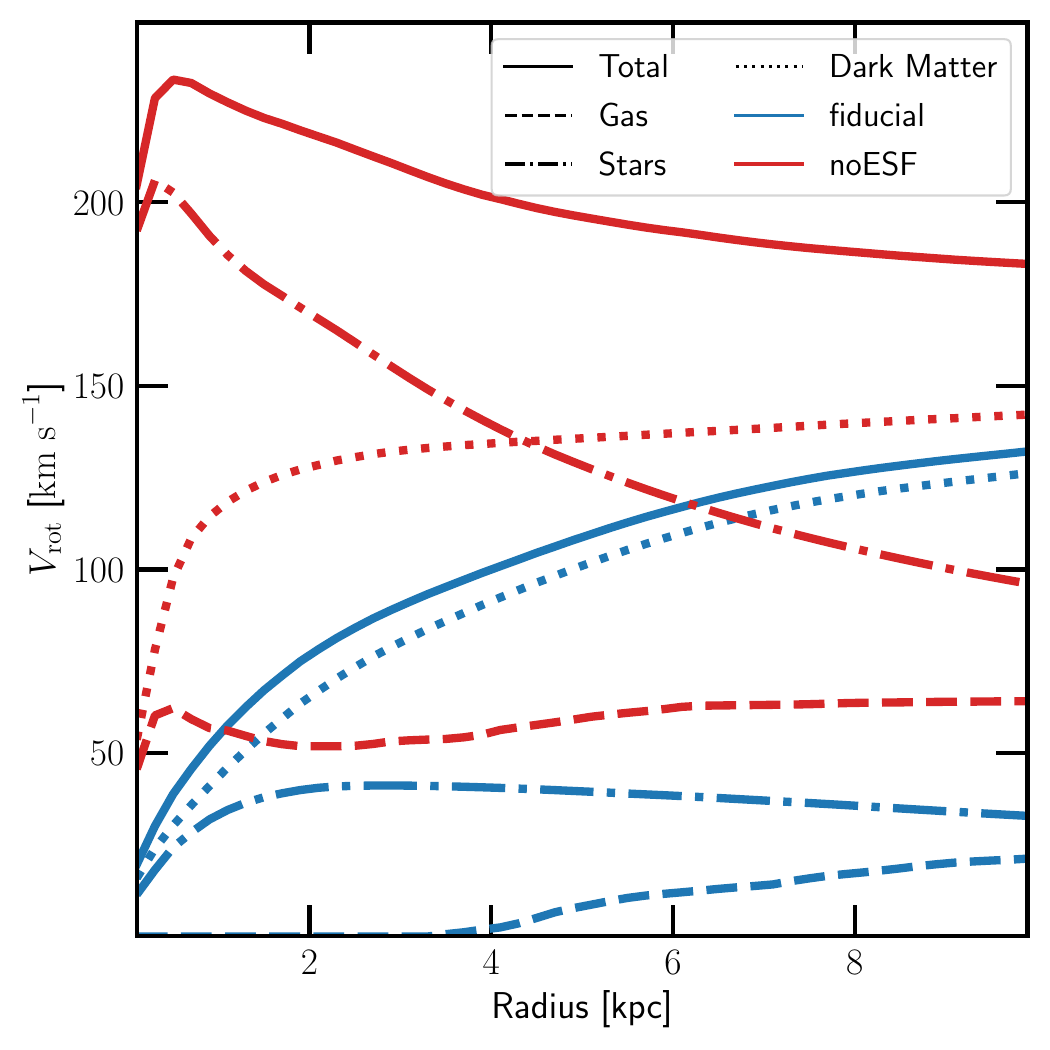}
    \caption{Rotation curves of the \target galaxies in the \mbox{m$11.5\_4\mathrm{x}$} (blue curves) and \mbox{m$11.5\_4\mathrm{x}\_\mathrm{noESF}$} (red curves) runs at $z=3$. The dotted, dashed, and dot-dashed lines are the contributions from dark matter, gas, and stars, respectively, while the solid line is the total rotation curve. Early stellar feedback helps reduce the central stellar mass density, which in turn produces flat, slowly rising rotation curves.}
    \label{fig:rotcurve}
\end{figure}

\begin{figure*}
	\includegraphics[width=0.49\textwidth]{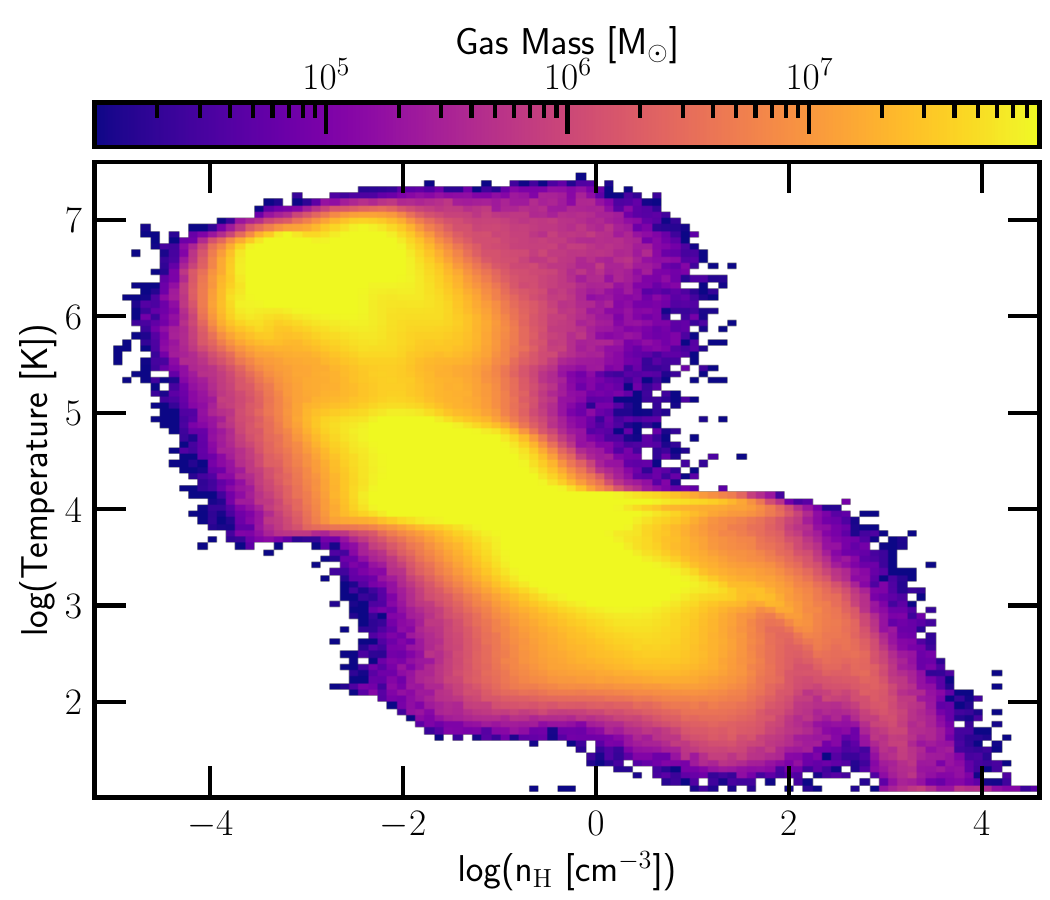}
 	\includegraphics[width=0.49\textwidth]{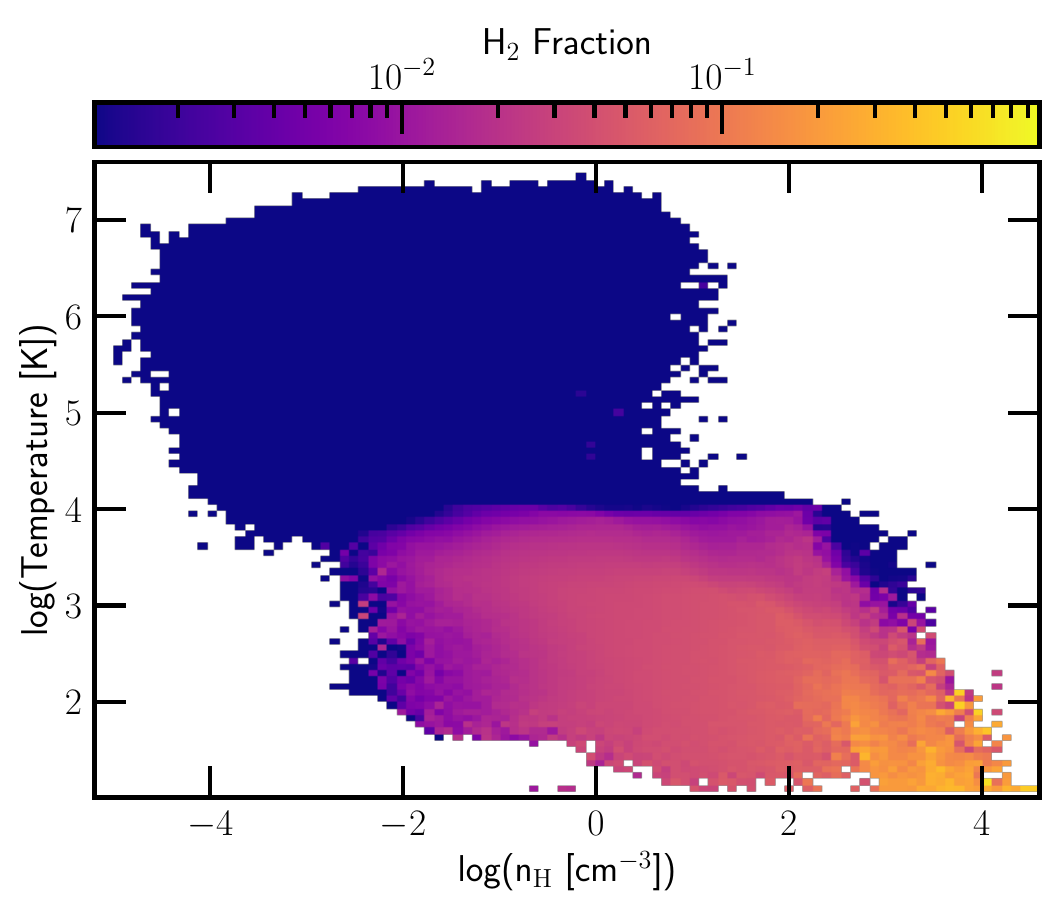}
   \includegraphics[width=0.49\textwidth]{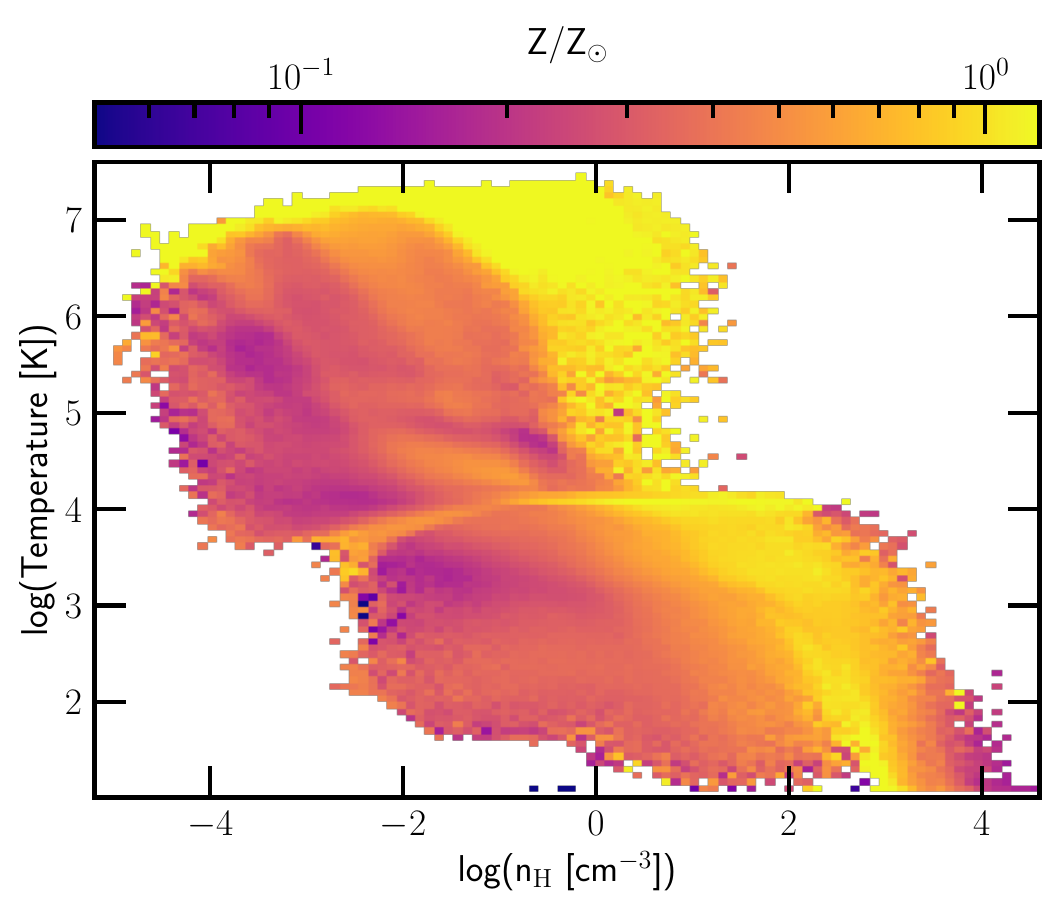}
   \includegraphics[width=0.49\textwidth]{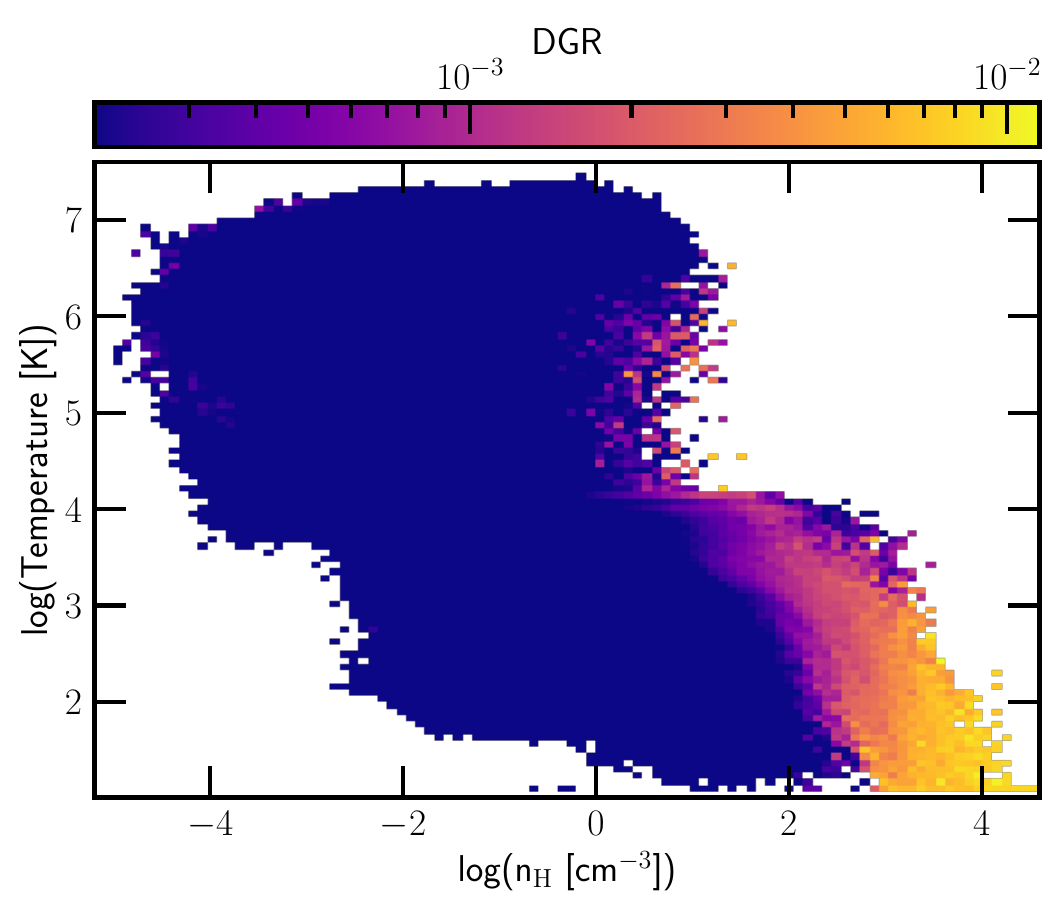}
    \caption{The temperature-density phase-space diagram at $z=3$, plotted for gas within the virial radius of the \target galaxy in the \mbox{m$12.6\_4\mathrm{x}$} run, weighted by the gas mass (top left panel), molecular hydrogen fraction (top right panel), gas phase metallicity normalized to the solar value (bottom left panel) and the dust-to-gas ratio (bottom right panel).}
    \label{fig:phasespace}
\end{figure*}

Therefore, many of the important processes that help regulate the star-formation rates in low-redshift galaxies become ineffective at higher redshifts \citep{Grudic2018, Grudic2019}. This can, in principle, lead to highly efficient star formation in the very early Universe ($z\gtrsim9$), which might explain the abundance of massive, bright galaxies seen by \jwst. This might also partially explain the tendency of some feedback models that have primarily been tested and/or developed in low-redshift environments to overproduce the stellar masses of galaxies at $3\lesssim z <9$ \citep{Feldmann2023, prfm}.

However, it is important to note that a number of physical processes have been largely overlooked in a majority of galaxy formation models. For example, radiation pressure from \lya scattering is capable of injecting up to $\sim100$ times more momentum into the ISM than UV continuum radiation pressure and stellar winds \citep{Nebrin2024}, particularly in the dust-poor environments of early galaxies. The low metallicity environments are also conducive for forming extremely massive stars \citep[$\sim100\,\Msun$; ][]{Schaerer2002} with a top-heavy IMF. This leads to a much larger radiation output ($\sim10\times$), more numerous SN per unit stellar mass formed, and a higher average SN explosion energy \citep{Heger2003, Wise2012} compared to low-redshift environments. These processes could potentially help suppress star formation in high-redshift environments. Unfortunately, the \thzoom simulations are not able to accurately capture the effect of these processes. 

Additionally, it should be noted that currently it is not possible to reliably resolve the Str\"omgren radius around newly formed stars. This limitation can lead to reduced photoheating, which in turn results in inaccurate momentum injection during these early stages. Although the modifications outlined in \citet{Deng2024a} help minimize the impact of unresolved \HII regions, their efficacy in realistic multiphase ISM environments is not well understood. Radiation pressure (primarily from UV photons) can also be underestimated in unresolved regions \citep{Hopkins2019}. This can be partially overcome by setting the direction of the photon flux radially outwards from the star particle as outlined in \citet{Kannan2020a}. However, our extensive tests showed that this did not have any significant impact on the total star-formation rate of the galaxy. This may be because, in high-density regions, the additional momentum from trapped IR radiation is necessary to disrupt the surrounding gas around newly formed stars \citep{Hopkins2012a}. Unfortunately, in high optical depth regions, the inherent numerical diffusion of the radiative transfer solver reduces the trapping factor of the IR photons leading to a lower than expected injection of momentum \citep{Rosdahl2015, Kannan2019}. Approximate methods that paint \HII regions around newly formed stars \citep{Smith2021} and/or compute IR trapping using a Sobolev approximation \citep{Hopkins2018, Hopkins2023} can potentially replicate these effects, helping reduce star-formation rates in high-redshift environments. However, ultimately, only simulations with significantly higher gas particle resolution than stellar resolution \citep{Peters2017}, or those resolving regions around young stars in a super-Lagrangian manner \citep{Hopkins2024} are capable of modeling these processes from first principles.

\begin{figure*}
	\includegraphics[width=0.99\textwidth]{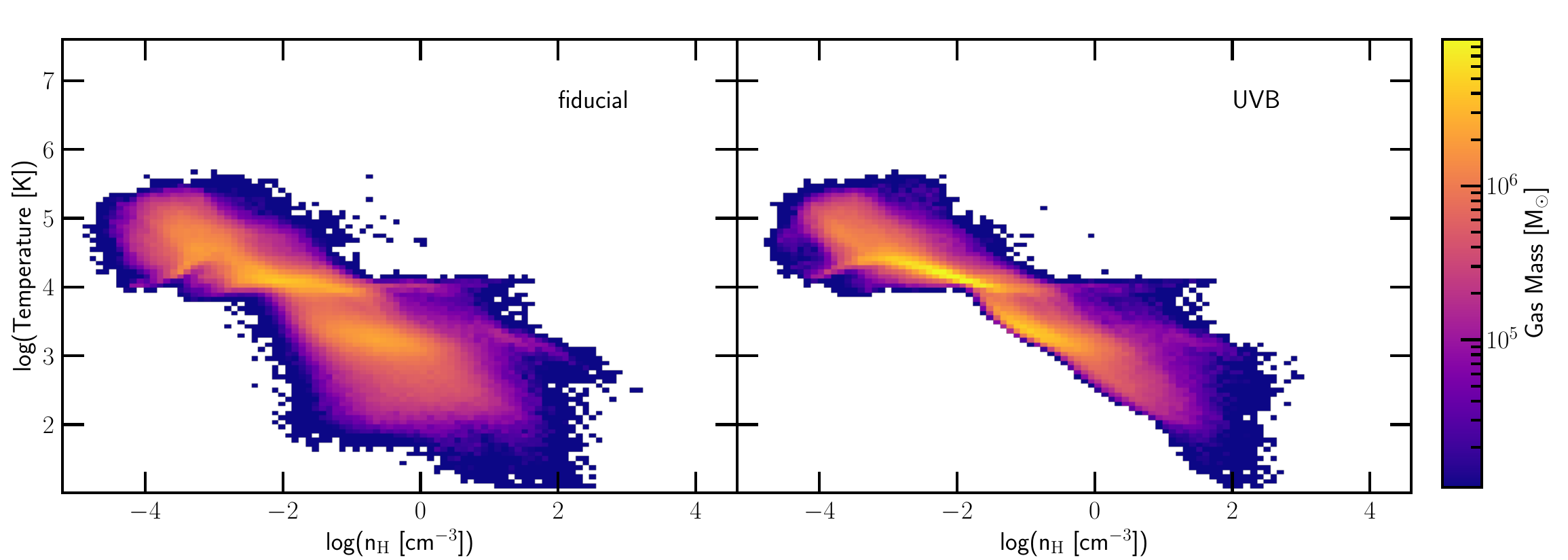}
    \caption{The temperature-density phase-space diagram (weighted by the gas mass) for gas within the virial radius of the \target galaxy in the \mbox{m$10.4\_8\mathrm{x}$} (left panel) and the \mbox{m$10.4\_8\mathrm{x}\_\mathrm{UVB}$} (right panel) runs at $z=3$. The amount of low-density, low-temperature gas is greatly enhanced when the widely used spatially constant UV background is replaced with a patchy radiation field taken from the parent \thesanone simulation.}
    \label{fig:nuvb}
\end{figure*}

While we plan to explore further improvements in the future, in the current simulations we resort to using an empirical model to emulate the impact of either additional missing physics and/or the inability to properly resolve/model feedback processes. The model injects momentum into neighboring gas cells of newly formed stars for a duration of $5~\mathrm{Myr}$ after the star particle is formed. This effectively disperses star forming gas clouds and suppresses the build-up of stars from low angular momentum gas in the centers of galaxies at early times. This is shown more clearly in Figure~\ref{fig:rotcurve}, which shows the rotation curve, which is the circular velocity $v_c = \sqrt{GM(<r)/r}$ as a function of radius for the \target galaxies in the \mbox{m$11.5\_4\mathrm{x}$} (red curves) and \mbox{m$11.5\_4\mathrm{x}\_\mathrm{noESF}$} (blue curves) runs at $z=3$. The solid curves show the total rotation curve, which is the sum of contributions from the dark matter (dotted curves), gas (dashed curves), and stellar (dot-dashed curves) components.  The \mbox{m$11.5\_4\mathrm{x}\_\mathrm{noESF}$} simulation has a centrally peaked rotation curve, while the fiducial simulation, \mbox{m$11.5\_4\mathrm{x}$},  has a slowly rising rotation curve which flattens out at large distances. This is mainly due to the fact that the noESF run has a large concentration of stars within the central kpc, while the stellar distribution is more extended in the fiducial model. In fact, the gravity from the central concentration of stars in the noESF run is able to enhance the DM and gas distribution in the center as well.   This plot clearly shows that the fiducial model does a good job, not only in regulating the stellar properties of galaxies but also helps produce flat, slowly rising rotation curves which are more commonly observed in the low-redshift Universe \citep{Sofue2001}.

We will note that a variety of different models, like reducing the minimum progenitor mass for SN explosion from the fiducial value of $8~\Msun$ to $6~\Msun$ (which doubles the number of SN per unit mass of stars formed) and boosting the amount of energy injected per SN were also tested in the context of regulating the stellar masses of simulated galaxies. However, they were unable to reduce the star formation efficiency, especially at high redshift. This is because the delay time of about $\sim 5~\mathrm{Myr}$ between the formation of a star particle and the first SN explosion allows most of the dense gas in the star-forming region to be converted into  stars. This leads to very clustered and extremely bursty star formation which in turn generates large mass and energy outflow rates \citep{Hopkins2020, Hu2023}. Despite this, due to the large reservoir of cold gas available at higher redshifts, within a dynamical time the gas content of the galaxy is replenished and forms stars again. Therefore, the ability of these delayed feedback models to regulate star formation was found to be deficient. ESF, on the other hand, starts injecting feedback energy as soon as a star particle is formed. This helps disperse the dense gas around the newly formed stars more efficiently, leading to a better regulation of the SFR in galaxies. We anticipate a follow-up paper including details about the numerical implementations and a more detailed discussion of the effects of the different feedback models. Finally, we will note that, recent observations of over-massive blackholes at high redshift \citep{Mezcua2024} might indicate rapid growth \citep{Pacucci2024} at early times, accompanied by significant amounts of feedback energy being injected into their surrounding environments. As black hole feedback is not included in our simulations, we cannot accurately estimate its effect on the galaxy population, particularly at the high-mass end. We aim to incorporate an appropriate black hole growth and feedback model in future works.

\subsection{Multi-phase ISM}

One of the main advantages of the \thzoom simulations compared to the parent \thesan volume is the inclusion of a state-of-the-art resolved multi-phase ISM model. \thesan uses an effective Equation of State model outlined in \citet{Springel2003} to model a multitude of important subgrid processes like gas cooling and heating due to SN feedback. Feedback is modeled by decoupling the ejected galactic wind particles from hydrodynamics to achieve better control over mass loadings and improve convergence \citep{Vogelsberger2013}. This model is numerically efficient and has been used to perform large-volume hydrodynamic simulations like Illustris, IllustrisTNG and MillenniumTNG \citep{Vogelsberger2014, Springel2018, KannanThesan, Pakmor2023, Kannan2023}. Although it has been quite successful in matching a wide array of integrated galaxy properties such as the galaxy luminosity function, color bi-modality, galaxy sizes, and metallicities, it is unable to predict the resolved structure ($\leq 1~\mathrm{kpc}$) of the multi-phase gas in the ISM of galaxies. The ISM model used in the \thzoom simulations has previously been successful in reproducing local ($z=0$) ISM observations like the dust-to-gas ratios \citep{Kannan2020b} and temperatures, accurate emission line luminosities like \lya, H$\alpha$, \citep{Smith2022b, Tacchella2022} and the nature of diffuse ionized gas in star-forming galaxies \citep{McClymont2024}. 

The ability of the \thzoom simulations to model the multi-phase ISM in a cosmological setup is shown in Figure~\ref{fig:phasespace}, which plots the 2-dimensional temperature-density histogram for the gas in the \target galaxy of the \mbox{m$12.6\_4\mathrm{x}$} run at $z=3$. The four different panels show this phase-space diagram weighted by the gas mass (top left panel), molecular hydrogen fraction (top right panel), gas phase metallicity normalized to the solar value ($Z_\odot=0.0127$; bottom left panel), and the dust-to-gas ratio (bottom right panel). The gas is distributed over a wide range of densities ($10^{-4}~\mathrm{cm}^{-3}\lesssim n_\mathrm{H} \lesssim 10^{4}~\mathrm{cm}^{-3}$) and temperatures ($10~\mathrm{K}\lesssim T \lesssim 10^{7}~\mathrm{K}$). The low-density gas is unable to cool below $\sim 10^4~\mathrm{K}$ due to heating from both the local and background radiation fields. Only the gas above a self-shielding density threshold of $n_\mathrm{H} \sim 10^{-2}~\mathrm{cm}^{-3}$ manages to cool below this threshold with temperatures reaching as low as $10-100$~K. The photoheated \ion{H}{II} regions are seen clearly as a build-up of high-density gas ($n_\mathrm{H} \gtrsim 10~\mathrm{cm}^{-3}$) at $\sim10^4$~K. Above $n_\mathrm{H} \gtrsim 10^3~\mathrm{cm}^{-3}$, the gas cooling rates are higher due to the higher abundance of molecular hydrogen. The high-density gas is self-shielded against the Lyman-Werner radiation and is dense enough to deposit most of the metals onto dust. The dust-to-gas ratios in this high-density gas reach values as high as $0.01$, which is the canonical MW value. These highly dusty environments are conducive for producing molecular \ion{H}{$_2$} as its formation involves catalytic reactions on the surface of interstellar grains. 

Feedback from newly formed stars ejects gas from the centers of galaxies into the circum-galactic medium (CGM). This gas  occupies the high temperature ($T\gtrsim10^5~\mathrm{K}$), relatively high-density ($n_\mathrm{H} \gtrsim 1~\mathrm{cm}^{-3}$) region of the phase-space as evidenced by its high metal content compared to other low-density but high-temperature ($T\gtrsim10^5~\mathrm{K}$) gas in the CGM. Almost all of this high-temperature gas is dust-free due to thermal sputtering. The metal-enriched outflows gradually mix into the surrounding CGM, enriching it with metals. The properties of CGM and its metal content will be discussed in detail in an accompanying paper (Pruto et al., in prep.). 

The phase-space structure of gas in galaxies is influenced by both the local radiation field from the stars in the galaxy and external large-scale radiation produced by other nearby galaxies. As outlined in section~\ref{sec:uvb}, the \thzoom simulations replace the commonly used time-varying but spatially constant UV background models with a novel model that accounts for the patchy and time-varying nature of the reionization process. Using radiation field maps from the parent \thesanone simulation, we follow the propagation of radiation across the boundary of the
high-resolution region to ensure incoming radiation flows present in the parent \thesan simulation are seen by the
high-resolution region of the zoom-in run. This is done until the final output of the parent simulation ($z=5.5$) before switching smoothly to the homogeneous UVB model from \citet{FG09}.

Figure~\ref{fig:nuvb} shows the difference in the phase-space structure of the gas in the \target galaxies of the  \mbox{m$10.4\_8\mathrm{x}$} (left panel) and \mbox{m$10.4\_8\mathrm{x}\_\mathrm{UVB}$} (right panel) runs at $z=3$. The most significant change is the lack of low density ($n_\mathrm{H} \lesssim 1~\mathrm{cm}^{-3}$) and low temperature ($T \lesssim 10^{4}~\mathrm{K}$) gas in the homogeneous UVB simulation. This is despite using the self-shielding prescriptions outlined in \citet{Rahmati2013} in the  \mbox{m$10.4\_8\mathrm{x}\_\mathrm{UVB}$} run. The fiducial simulation is able to retain this low-density, low-temperature gas in the galaxy even at $z=3$, long after we have switched to injecting the uniform UV background at the boundary of the high-resolution region.  This is because the self-shielding prescriptions are purely local, meaning that a parcel of gas is shielded against radiation based on its density. However, when the UV background is handled by radiative transfer, dense clouds in galaxies can cast shadows behind them which allows the gas to remain cool and neutral despite having a low density. A more thorough investigation into the effect of the UV background on the properties of galaxies is outlined in an accompanying paper \citep{Zier2025}.

\begin{figure}
	\includegraphics[width=0.99\columnwidth]{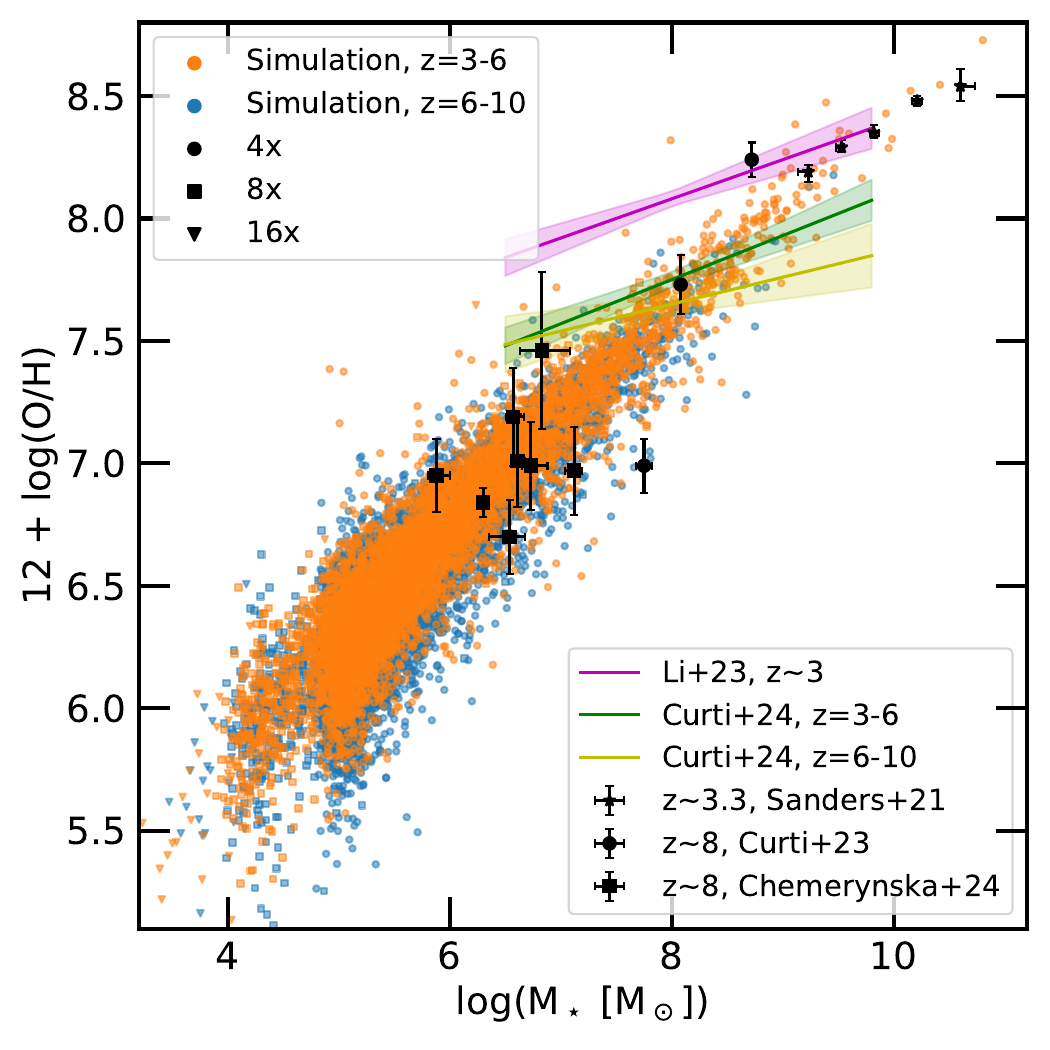}
    \caption{The stellar mass (within twice the stellar half mass radius) -gas phase metallicity (within the stellar half mass radius) relation for all the \central galaxies in the simulation suite, grouped into two redshift bins (z=3-6 and z=6-10), compared to recent HST \citep{Sanders2021} and \jwst \citep{Curti2023, Li2023, Curti2024, Chem2024} observations. There is very little redshift evolution and resolution dependence of the mass metallicity relation in the \thzoom simulations.}
    \label{fig:metals}
\end{figure}

\subsection{Metal and dust enrichment}
\label{sec:metals}

The metal and dust content of galaxies play a crucial role in their evolution. They are responsible for a large fraction of radiative cooling in the gas, which in turn influences gas accretion and star-formation rates. Additionally,  they regulate the attenuation and escape of radiation from the host galaxy \citep{Smith2022b}. It is, therefore, important for simulations to properly model these quantities to obtain accurate galaxy properties. Figure~\ref{fig:metals} plots the average gas phase metallicity within the stellar half-mass radius as a function of stellar mass (within twice the stellar half-mass radius) for all the \central galaxies in the \thzoom simulation suite. The results for the \fx, \ex, and \sx resolution levels are shown as circles, squares, and inverted triangles, respectively. All three resolutions show similar metal enrichment in the mass range they overlap, indicating good convergence. Simulated galaxies are grouped into two redshift regimes, $z=3-6$ (orange points) and $z=6-10$ (blue points). The gas metallicity is expressed in terms of $12 + \mathrm{log(O/H)}$, which is calculated by assuming that the solar metallicity in these units is 8.69 \citep{Asplund2009}. For comparison, we plot the $z\sim8$ observational estimates from \jwst \citep[black points and squares; ][]{Curti2023, Chem2024} and the estimates from HST (black stars) for the metallicity of galaxies at cosmic noon \citep[$z\sim3.3$; ][]{Sanders2021}. We also plot the fit to the observational data provided by \citet{Li2023} at $z\sim3$ (pink line) and the fits at $z=3-6$ (green line) and $z=6-10$ (yellow line) outlined in \citet{Curti2024}. The simulations do a good job of reproducing the observed metallicity of galaxies over the wide redshift range. However, it must be noted that the observational fits have a shallower slope compared to the simulation results, which matches other simulation results that model high redshift galaxies \citep[see for e.g.,][]{Wilkins2023, Marszewski2024}. This is possibly due to the difficulty of observationally measuring the metallicity of less metal-enriched galaxies. A more detailed investigation into the the metal abundance of simulated galaxies is presented in accompanying work \citep{McClymont2025c}.

\begin{figure}
	\includegraphics[width=0.99\columnwidth]{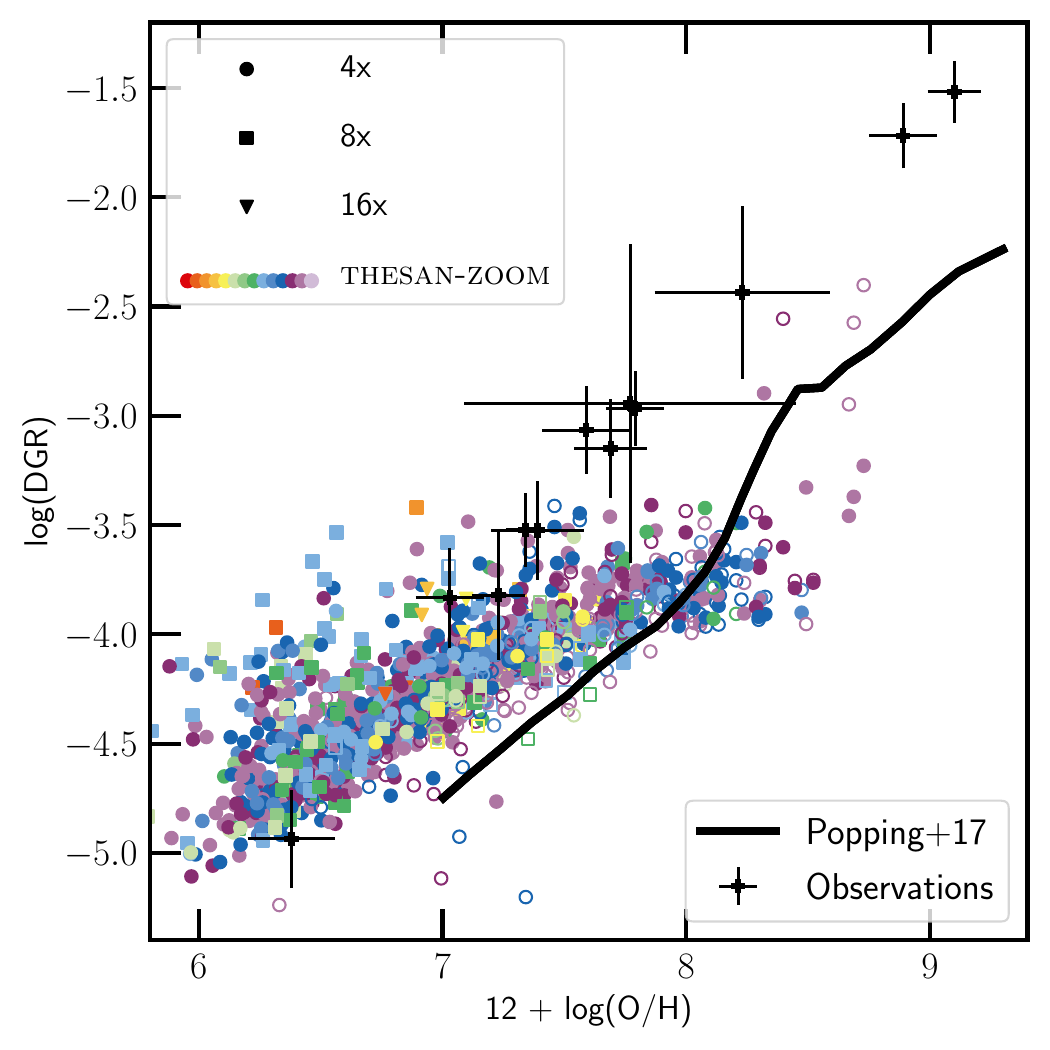}
    \caption{Dust-to-gas ratio (DGR) - gas metallicity (within the stellar half mass radius) relation at $z=3-3.5$ for all the \central galaxies in the simulation suite, with the colors indicating the different runs as outlined in Figure~\ref{fig:smhm}. The solid-filled points are the DGRs calculated within the stellar half-mass radius, while the open points are DGRs calculated only for gas eligible for star formation. High-density star-forming gas is much more dust enriched than the mean ISM. The black line shows the result from the empirical dust models of \citet{Popping2017} while the black points are observational estimates from \citet{DeCia2016} and \citet{Wiseman2017}. While the simulations do a good job of reproducing the dust-to-gas ratios in low metallicity environments, it remains challenging to get a significant build-up of dust in high-metallicity galaxies.}
    \label{fig:dust}
\end{figure}

Figure~\ref{fig:dust} plots the dust-to-gas ratio (DGR) as a function of the gas phase metallicity (within the stellar half mass radius) for all the \central galaxies in the \thzoom simulation suite. The different colors and symbols indicate different resolutions and simulations. The solid-filled points are the DGRs calculated within the stellar half-mass radius, while the open points are DGRs calculated for gas cells eligible for star formation. The high-density star-forming regions are generally more dusty than the rest of the ISM especially in high-mass galaxies. This is due to the increased rate of deposition of gas-phase metals onto existing dust grains at higher densities and metallicities. The solid black curve depicts the relation derived from the semi-analytical (SAM) dust model outlined in \citet{Popping2017}. The black points are the estimates of the DGRs derived from observations of damped \lya absorbers (DLAs) seen in the spectra of high-redshift quasars \citep{DeCia2016} and gamma-ray bursts \citep[GRB-DLAs;][]{Wiseman2017}. In the low metallicity regime, the dust content of the simulated galaxies is slightly higher than the estimates from the SAM but aligns well with observational estimates. At higher metallicities ($12 + \mathrm{log(O/H)} \gtrsim 8$), there is a notable increase in the DGRs, mirroring the model proposed by \citet{Popping2017}. This increase is particularly evident when considering the dust content in high-density star-forming gas. Conversely, the dust enrichment in the average ISM gas shows only a modest increase with rising metallicity. The significant rise in dust content in high-density and high-metallicity gas is a result of modifications to the metal deposition rates, which are now dependent on both density and metallicity (see Section~\ref{sec:dust} for more details). This behavior was absent in the original \thesan simulations because it did not account for the additional dependence on the metallicity of the gas when calculating dust growth rates in the ISM.

\begin{figure}
	\includegraphics[width=0.99\columnwidth]{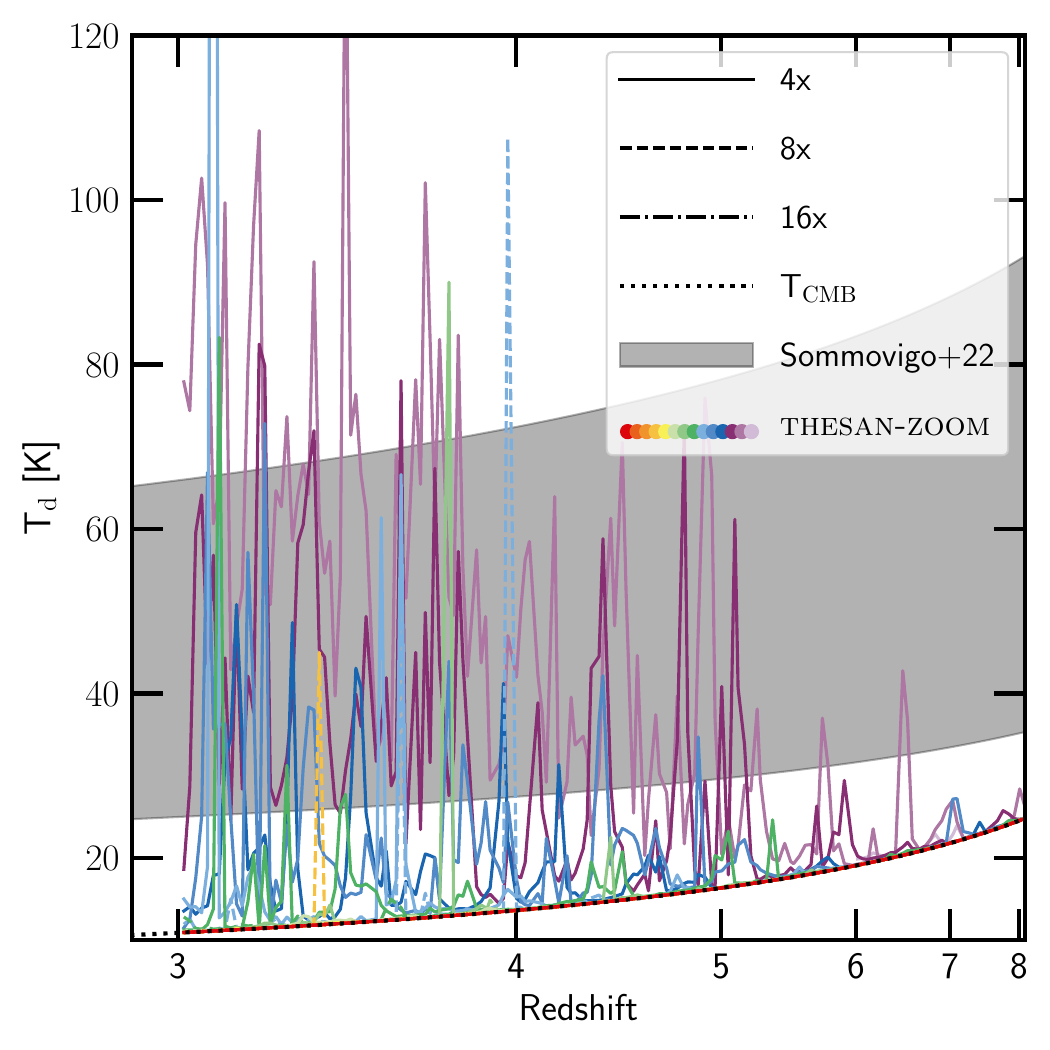}
    \caption{Luminosity weighted dust temperature as a function of redshift for the \target galaxies, with the colors indicating the different simulations as outlined in Figure~\ref{fig:smhm}. The different line styles indicate the different resolution levels: namely, $4x$, $8x$, and $16x$, by solid, dashed and, dot-dashed lines, respectively. The grey region plots the expected dust temperatures from the physical model outlined in \citet{Sommovigo2022}. In our model, the dust temperature is highly variable, with the peaks in the dust temperature corresponding to an increase
in the local radiation field intensity associated with bursts of
star formation in the galaxy.}
    \label{fig:dustT}
\end{figure}

Finally, Figure~\ref{fig:dustT} plots the evolution of the luminosity-weighted dust temperature for all the \target galaxies in the \thzoom simulations as a function of redshift. The colored solid, dashed, and dotted lines denote the dust temperatures in the \fx, \ex, and \sx resolution levels, respectively. The dotted black curve is the CMB temperature, and the gray shaded area is the physical model outlined in \citet{Sommovigo2022}, which attributes the increasing dust temperature with redshift to the decrease in gas depletion times.

The simulated dust temperatures seem to fluctuate on very short timescales with the minimum dust temperature being set by the CMB. This is because the dust in galaxies is in radiative equilibrium with the radiation field \citep{Draine2003}. CMB provides the underlying background radiation field intensity, while the far-UV and visible light from stars that are absorbed by dust and re-radiated in the IR provide the rest of the radiation intensity. Moreover, dust–gas collisions lead to energy exchange, resulting in an additional cooling/heating mechanism of dust \citep{Burke1983}. The dust temperature is then calculated by solving the instantaneous equilibrium condition between the radiative and collisional cooling rates (see Section~\ref{sec:dust} for more details). Therefore, the peaks in the dust temperature correspond to an increase in the local radiation field intensity associated with bursts of star formation in the galaxy. The maximum dust temperatures mostly lie within the temperature range predicted by the model outlined in \citet{Sommovigo2022}, but at lower redshifts, massive starbursts can produce very high dust temperatures of the order $100$~K. It is quite clear that the \thzoom simulations do a relatively good job of predicting the right metal and dust enrichment of high-redshift galaxies and are capable of reproducing sensible dust temperatures, although building up significant dust in high-metallicity environments remains a challenge. A more thorough investigation into the properties of dust in galaxies is shown in an accompanying paper (Garaldi et al., in prep.).

\subsection{Statistical properties of galaxies}

\begin{figure}
	\includegraphics[width=0.99\columnwidth]{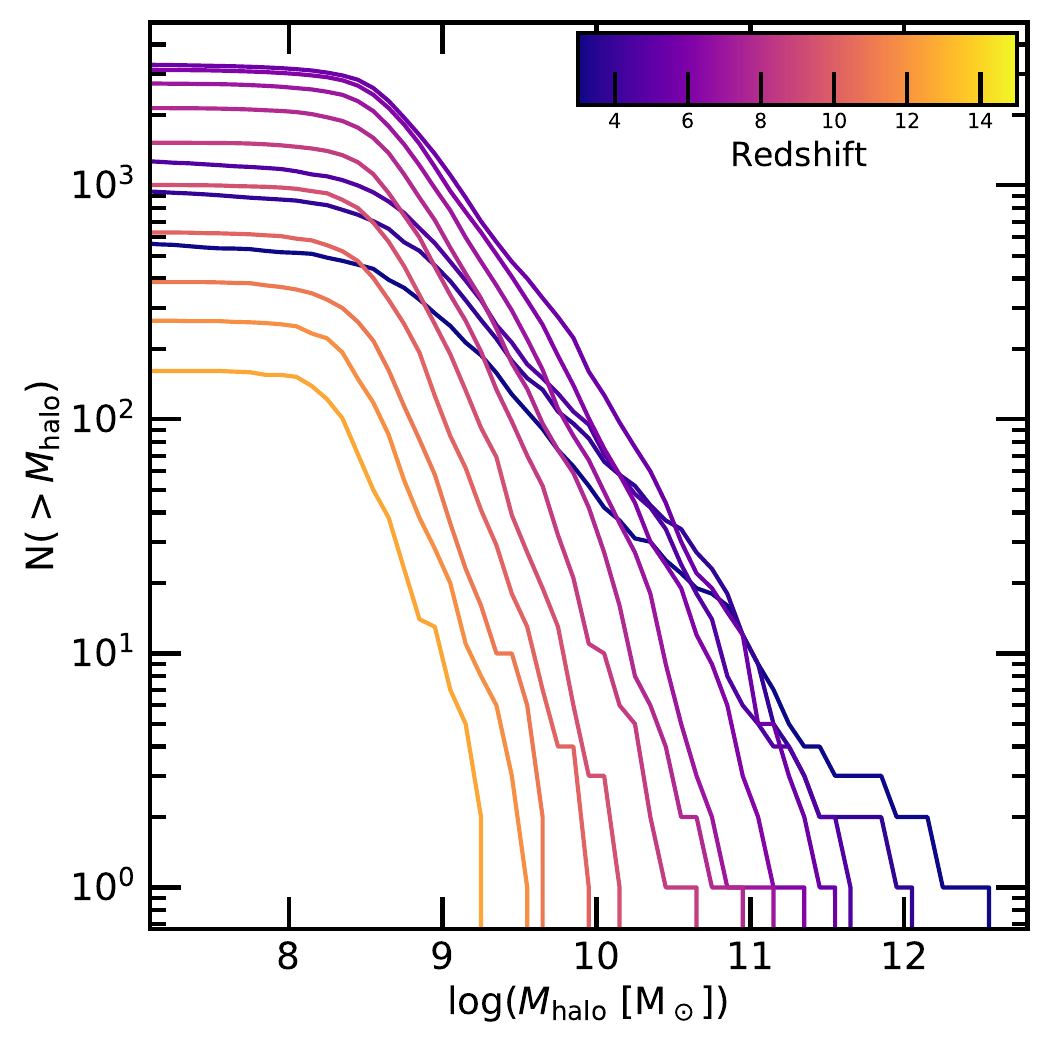}
    \caption{The cumulative number distribution of the  \central galaxies in the \thzoom simulations as a function of their halo mass. Only well-resolved halos that contain at least 10 stellar particles are considered in this work. The different colored lines show the distribution at $z=3-15$ as indicated. As expected, there are many more smaller galaxies in the simulation sample, while the highest-mass halos are more sparsely sampled.}
    \label{fig:mh}
\end{figure}

The simulations presented in this work, by virtue of it being $14$ individual zoom-in regions, do not contain information about the statistical properties of galaxies such as the halo mass functions, stellar mass functions, or UV luminosity functions. However, it is possible to estimate these quantities by correcting for the biased selection function of the zoom-in halos. We do this by using a variation of the method outlined in \citet{Ma2018}, which assigns weights to the simulated halos to recover the appropriate number density of halos in the Universe (see also \citealt{Sun2023}). For each of the $188$ snapshots, the \central galaxies in all the fiducial simulations (including different resolutions) are binned according to their halo mass. We consider all galaxies in the mass range $7 \leq \mathrm{log}(M_\mathrm{halo} [\Msun]) \leq 13$, that host at least $10$ star particles within their virial radius. We use a bin width of $\Delta \mathrm{log}M_\mathrm{halo}$ of $0.1$, although the exact value does not affect our results as long as it is not too large ($\lesssim0.5$). Figure~\ref{fig:mh} shows the cumulative number distribution of the \central galaxies as a function of their halo mass at $z=3-15$, in increments of $\Delta z =1$, as indicated. The total number of galaxies at a particular redshift increases from about $100$ at $z=15$ to $\sim 5000$ at $z=6$ and back to $\sim 1000$ at z=3. The reduction in the number of galaxies at later times is due to the fact that the \mbox{m$13.0\_4\mathrm{x}$} simulation is only run down to $z=6$. Being the most massive halo in the simulation suite, it has the largest high-resolution Lagrangian region, leading to a large number of lower-mass halos being resolved in addition to the \target galaxy of the simulation. As expected, there are many more smaller galaxies in the simulation sample, while the highest-mass halos are sampled only by a handful of halos.

\begin{figure*}
	\includegraphics[width=0.98\textwidth]{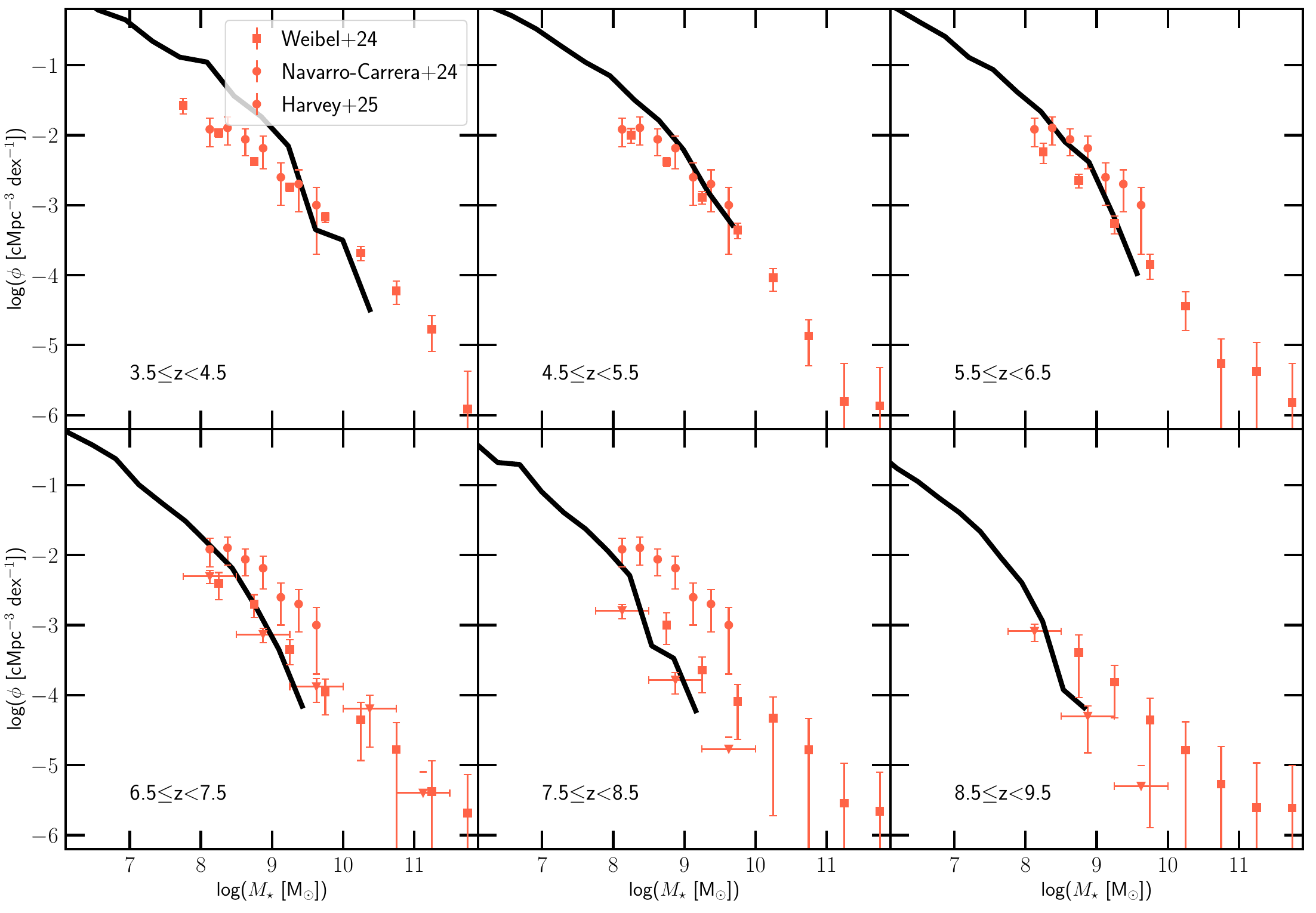}
    \caption{The galaxy stellar mass (within twice the stellar half mass radius) function estimated from the \thzoom simulations at $z=3.5-9.5$ as labeled. For comparison we show recent observational estimates from deep \jwst data by \citet[red circles; ][]{NC2024},  \citet[red squares; ][]{Weibel2024}, and \citet[red inverted tiangles; ][]{Harvey2025}. Schechter function fits to the simulated GSMFs are tabulated in Table~\ref{table:gsmf} of Appendix~\ref{ap:ap1}. }
    \label{fig:gsmf}
\end{figure*}

The expected number of halos in the Universe is estimated using the code \textsc{HMFcalc} \citep{hmf}. The halo mass functions are estimated by employing the same cosmological parameters and the definition of virial overdensities used in the \thesan simulations. Specifically, we use the fitting functions from \citet{Behroozi2013}, which is a modified version of the \citet{Tinker2008} results. We note that these mass functions match very well with the ones extracted from our large-volume \thesanone 
box in the redshift range we consider here \citep{Smith2022}. These mass functions are calculated for every redshift corresponding to each of the $188$ snapshot outputs. We then bin the mass function in $\Delta \mathrm{log}M_\mathrm{halo}$ of $0.1$ and calculate the number of halos expected per mass bin ($N_\mathrm{expected}$) assuming a volume of $95.5~\mathrm{cMpc}^3$, which is the volume of the \thesan simulations. All simulated galaxies in the same bin are assigned the same weight $w=N_\mathrm{expected}/N_\mathrm{sim}$, where $N_\mathrm{sim}$ is the number of simulated galaxies in that bin, calculated using the method outlined in the previous paragraph. The halo mass/stellar mass/UV luminosity functions are then calculated by binning these quantities in the simulated range, summing the weights $w_i$ over all simulated halos contributing to that bin ($i$), and dividing by the corresponding volume. It must be noted that this method should recover the correct mass and luminosity functions as long as the simulated galaxy sample is not strongly biased. Our sample could possibly be biased due to the fact that most of the low-mass halos live within a few virial radii of the larger nearby \target halo. However, as long as the star formation efficiency is independent of the galaxy environment, the results presented using this scheme should be quite robust.

\begin{figure*}
	\includegraphics[width=0.98\textwidth]{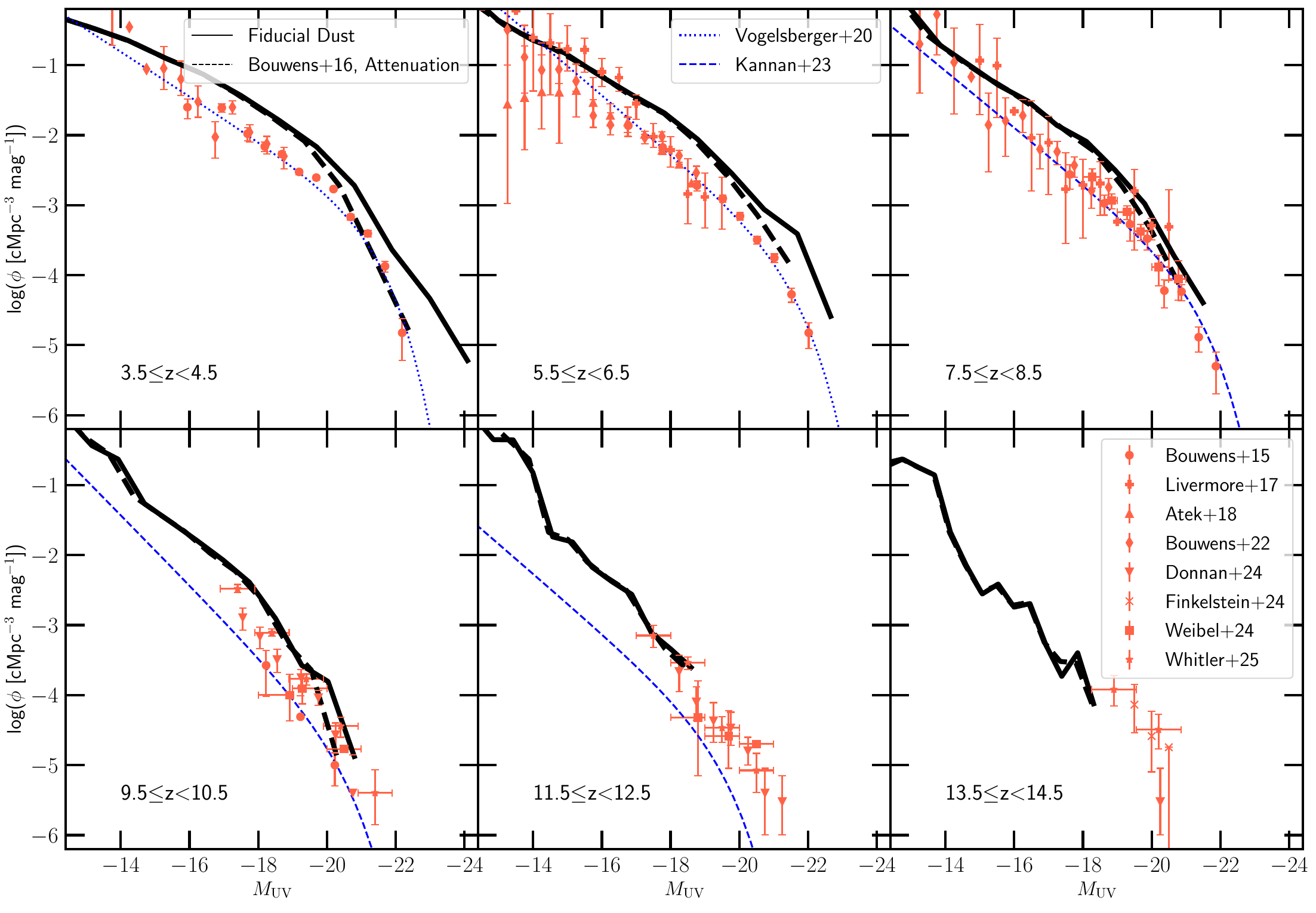}
    \caption{The UV luminosity function at rest frame $1500$~\AA~ from $z=3.5-14.5$. The solid black curve includes attenuation from the fiducial dust model of \thzoom, which overshoots the bright end at $z\lesssim7$ due to too little dust. Meanwhile the dashed black curve assumes a calibrated empirical dust attenuation relation presented in \citet{Bouwens2016}. Estimates from IllustrisTNG \citep{Vogelsberger2020b} and MillenniumTNG \citep{Kannan2023} are indicated by the dotted and dashed blue curves. Observational estimates using \textit{HST} and \jwst from \citet[red circles][]{Bouwens2015}, \citet[red pluses; ][]{Livermore2017}, \citet[red triangles; ][]{Atek2018}, \citet[red diamonds; ][]{Bouwens2022}, \citet[inverted red triangles; ][]{Donnan2024}, \citet[red crosses; ][]{Finkelstein2024} and \citet[red squares; ][]{Weibel2024} are also shown for comparison. Schechter function fits to the simulated UV luminosity functions are tabulated in Table~\ref{table:uvlf} of Appendix~\ref{ap:ap1}.}
    \label{fig:uvlf}
\end{figure*}

For a particular redshift $z$, the mass/luminosity functions are calculated by combining all the snapshots that fall within $z-0.5$ and $z+0.5$. By doing so, we sample the same halos multiple times and treat them as statistically equal in our analysis. This helps improve the statistics by increasing the number of halos contributing to a particular mass/luminosity bin. This is especially useful at the high-mass/luminosity end, which contains only a handful of halos in each bin per snapshot. This also helps account for the halo-to-halo variance and time variability of the galaxy properties.

Figure~\ref{fig:gsmf} shows the estimated galaxy stellar mass functions (solid black curves) from the \thzoom simulations using the method described in the previous paragraphs. We plot the mass functions for $z=4-9$ in intervals of $\Delta z=1$ as indicated. The red points are observational estimates from recent \jwst data. At $z=4$, the simulated stellar mass functions match the observational estimates at the high-mass end ($M_\star > 10^9~\Msun$) but overestimate the abundance of low stellar mass galaxies by about $0.2-0.4~\mathrm{dex}$. This seems to indicate a slightly higher than expected star-formation efficiency in the simulated low-mass halos.  From $z=5-7$, the simulations match the observations quite well in the mass range they overlap. At even higher redshifts, the \thzoom simulations seem to underpredict the number density of galaxies with $M_\star > 10^9~\Msun$, consistent with the over-abundance of massive galaxies at high-redshift reported by most recent works. It must, however, be noted that it is challenging to estimate the stellar masses due to the  uncertainties in the galaxy spectral energy distribution (SED) templates used for photometric fitting \citep{Endsley2023, Steinhardt2023}. This leads to number density estimates that are different by about $0.5~\mathrm{dex}$ between the observational estimates of \citet[red circles;][]{NC2024}, \citet[red squares;][]{Weibel2024}, and \citet[red inverted triangles; ][]{Harvey2025} shown in Figure~\ref{fig:gsmf}.

A more direct comparison with observations can be made by considering the UV luminosity functions at $z=4-14$ in intervals of $\Delta z =2$ as shown in Fig.~\ref{fig:uvlf}. The UV magnitude of each \central galaxy is obtained by summing up the radiation output at rest-frame $1500~\text{\AA}$ \citep[using BPASS version 2.2.1 tables;][]{Eldridge2017} of all the stars in the galaxy.  Due to the probabilistic nature of the star formation routine, the star-formation history will only be sparsely sampled in halos with low SFR. These halos will have long periods with zero star formation interspersed with sudden jumps in star formation as a new particle is stochastically spawned. This young massive star will dominate the entire radiation output of the galaxy, especially if the mass of the galaxy is close to the resolution limit of the simulation, which will adversely affect the UV luminosity functions. To overcome this numerical artifact, the age and mass of stars formed less than 5 Myr ago are smoothed over a timescale given by $t_\mathrm{smooth} = M_\star(< 5~\mathrm{Myr})/\mathrm{SFRgal}$, where $\text{SFR}_\text{gal}$ is the instantaneous SFR of the corresponding galaxy calculated by summing up all the SF probabilities of the cells eligible for star formation. This smoothing procedure is only done for halos with $t_\mathrm{smooth} > 5~\mathrm{Myr}$. We note that this only affects halos close to the resolution limit and allows for a more faithful prediction of the simulated UV luminosity function \citep{KannanThesan}.
 
The amount of dust attenuation is calculated using the post-processing Monte Carlo radiative transfer code \textsc{colt} \citep{Smith2015, Smith2022b}. Briefly, photon packets are launched from star particles and propagated out to the virial radius of the galaxy. The photons interact with the dust via scattering and absorptions, with the frequency-dependent albedos and opacities obtained from \citet{Weingartner2001}. The dust distribution and their properties, such as the fraction of carbonates and silicates, are taken directly from the simulation output. As PAHs are not directly tracked, we assume that $1\%$ of the total dust is in the form of PAHs, and we obtain this by subtracting the corresponding amount from the amount of carbonaceous dust predicted in the simulation. The black curves show the dust-attenuated UVLFs from the \thzoom simulations. For comparison, we show observational estimates (red points) from both \textit{HST} \citep{Bouwens2015, Livermore2017, Atek2018, Bouwens2022} and \jwst \citep{Donnan2024, Finkelstein2024, Weibel2024} observations. The dotted and dashed blue lines show the UVLF estimates from IllustrisTNG \citep{Vogelsberger2020} and MilleniumTNG \citep{Kannan2023} simulations, respectively. 

At $z=4$, the simulated luminosity function lies slightly above the observational values and the IllustrisTNG results for almost the entire range except in low-mass galaxies with $M_\mathrm{UV}>-17$. The inconsistency at intermediate luminosities mirrors the higher abundances of corresponding low mass galaxies seen in the simulated galaxy stellar mass function (Figure~\ref{fig:gsmf}). The discrepancy at the high luminosity end, on the other hand, is mainly due to the inability of our dust model to attenuate enough UV light from massive galaxies. This conclusion is backed up by the fact that using an empirically derived dust attenuation relation obtained by fitting the IRX–UV relationship inferred from ALMA observations at $z\sim 4-7$ \citep{Bouwens2016}, reproduces the correct number density of galaxies with $M_\mathrm{UV}<-19$ (dashed black curves). 

\begin{figure}
	\includegraphics[width=0.99\columnwidth]{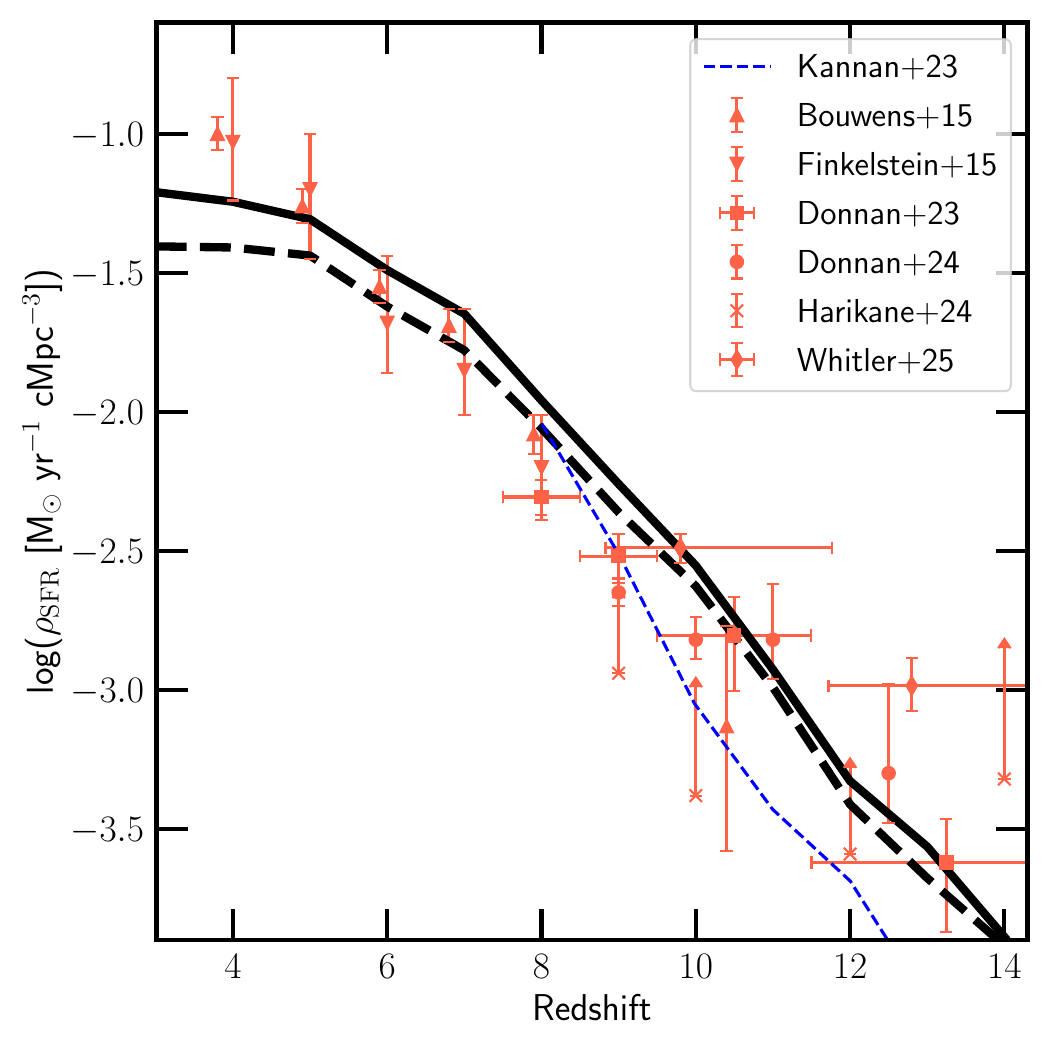}
    \caption{The evolution of the star-formation rate density (integrated down to M$_\mathrm{UV}=-17$) as a function of redshift from $z=3-14$. The solid black curves include attenuation from the fiducial dust model of \thzoom, while the dashed black curve assumes the  dust attenuation relation presented in \citet{Bouwens2016}. The dashed blue curve shows the SFRD estimated from MillenniumTNG \citep{Kannan2023}. The red symbols are observational estimates from \citet{Bouwens2015}, \citet{Finkelstein2015}, \citet{Donnan2023, Donnan2024} and \citet{Harikane2024} as indicated. This excellent match with the observations indicate that simulations do a good job of reproducing the
star-formation rates of high-redshift galaxies.}
    \label{fig:sfrd}
\end{figure}

However, we need to reconcile the fact that the dust-to-gas ratios of the simulated galaxies seem to agree well with expectations but do not recover the correct attenuation in high-mass galaxies. Our initial analysis shows that the dust mass of the simulated galaxies varies on very short time scales. There are periods of large dust accumulation as gas cools to high densities and forms stars, followed by feedback from stars heating up and driving large-scale outflows, which destroys most of the dust in the galaxy, leading to very low attenuation values. The time galaxies spend in the dusty phase seems too low to produce enough attenuation to match the observed UVLFs. We note that similar issues were also identified in simulations using \textsc{fire-2} with an enhanced dust species evolution model \citep{Choban2022}, which was attributed to a combination of low galactic metallicities and extremely bursty star formation \citep{Choban2025}.

Therefore, the empirical dust model is required to match the observed UVLFs, especially at lower redshifts ($z\lesssim8$). Beyond that ($z=10-14$), the dust attenuation is minimal, so the results from the fiducial dust model match those derived using the empirical dust model. At $z=10$,  the simulations seem to match the observed values in the luminosity overlap.  At even higher redshifts, the \thzoom simulations only contain galaxies with magnitudes extending down to $M_\mathrm{UV}\gtrsim-18$, whereas most of the observations are of galaxies with $M_\mathrm{UV}\lesssim-18$. While it does look like extending the simulation results by using double power-law or Schechter function fits might be able to reproduce the observed results, even higher mass galaxies taken from larger boxes may be necessary to probe this regime properly.  

Finally, Fig.~\ref{fig:sfrd} shows the evolution of the cosmic star-formation rate density as a function of redshift, which is derived by integrating the UVLFs down to $M_\mathrm{UV}=-17$. To be consistent with observations, we convert the UV luminosity density into a star-formation rate density by using the relation
\begin{equation}
    \mathrm{SFR~[\Msun~yr^{-1}]} = \kappa_\mathrm{UV}L_\mathrm{UV}~[\mathrm{erg~s^{-1}~Hz^{-1}}], 
\end{equation}
where $\kappa_\mathrm{UV}$ is the conversion factor with a value of $\kappa_\mathrm{UV} = 1.15 \times 10^{-28}~\mathrm{\Msun~yr^{-1}}/(\mathrm{erg~s^{-1}~Hz^{-1}})$,  which is valid for a \citet{Salpeter1955} IMF and consistent with the cosmic star-formation history out to $z\sim8$ \citep{Madau2014}. The solid black curve shows the result for UVLFs calculated using the fiducial dust model, while the dashed black curve is the estimate using the empirical dust model. We will note we can only integrate the UVLFs up to the highest luminosity galaxy present in the simulation sample because we are integrating the estimated UVLFs directly and not the functional fit to the data. This means that the estimated SFRD is a lower limit, especially at high redshifts where the most luminous galaxies in the simulated sample have $M_\mathrm{UV}\gtrsim-18$.  Observational estimates from \textit{HST} \citep{Bouwens2015, Finkelstein2015} and \jwst \citep{Donnan2023, Donnan2024, Harikane2024, Whitler2025} are denoted by red symbols. The dashed blue curve is the SFRD estimated from the MillenniumTNG simulations \citep{Kannan2023}.  The simulated SFRD matches very well with the observational estimates at all the redshifts considered, $z=3-14$, and is higher than the MillenniumTNG results by about $0.5~\mathrm{dex}$ at $z>8$. This excellent match with the observations tells us that the simulated sample does a good job of reproducing the star-formation rates of luminous galaxies at almost all redshifts considered. 

\section{Discussion and Conclusions}
\label{sec:conclusions}
We have introduced the \thzoom project, a suite of high-resolution zoom-in cosmological simulations of $14$ high-redshift ($z>3$) galaxies selected from the parent \thesan simulation volume \citep{KannanThesan}. They cover a wide range of halo masses from $M_\mathrm{halo} = 10^8 - 10^{13}~\Msun$ at $z=3$, and are simulated at three different resolution levels. In the highest-resolution runs, the DM and baryonic mass resolutions reach $762~\Msun$ and $142~\Msun$, respectively, corresponding to spatial resolutions of $\sim 140~\mathrm{cpc}$ for DM and stars, and down to $\sim 17~\mathrm{cpc}$ for gas. We emphasize that these numbers correspond to the gravitational softening length, so the smallest gas cells can be much smaller, such that the radiation and other physics calculations are tracked at higher resolutions. Due to the computational cost, we only simulate the most massive halos at the standard resolution level (4x), halos up to $\mathrm{M}_\mathrm{halo} \sim 10^{11} ~ \Msun$ at increased resolution (8x), and $\mathrm{M}_\mathrm{halo} \lesssim 10^{10} ~ \Msun$ at the highest resolution (16x).

Our galaxy formation model captures the multi-phase nature of the interstellar medium by self-consistently including the effects of supernova feedback, radiation fields, dust physics, and low temperature cooling via molecular hydrogen \citep{Kannan2020b, Kannan2021}. In addition to the fiducial simulations, we have performed a set of complementary variants designed to investigate the effect of changing the numerical and physical parameters on galaxy properties. In total, this gives us a set of $60$ simulations, which are used to investigate various properties of high-redshift galaxies. In this first paper, we focus on a broad characterization of the simulated galaxies, including investigating the stellar content and star-formation rates, structure of the ISM, metal and dust enrichment, and the large-scale statistical properties of the simulated sample. We compare these results with observations from \hst, \jwst, and other telescopes where possible. The main findings are summarized below:
\begin{enumerate}
    \item The high-mass end of the simulated stellar-to-halo mass relation, specifically for $M_\mathrm{200c}\gtrsim 10^{10}~\Msun$ is consistent with the abundance matching estimates from from \citet{Moster2018}, \citet{Tacchella2018} and \citet{Behroozi2019}. In contrast, low-mass galaxies appear to overproduce stellar masses by approximately $\sim0.2-0.4~\mathrm{dex}$, particularly at low redshifts, $z=3-5$.
    
    \item The simulated star-forming main sequence shows a much stronger evolution compared to the fits to the observed main sequence outlined in \citet{Popesso2023}. At $z\geq10$, almost all the simulated galaxies show increasing star formation histories. At later times higher-mass galaxies tend to have close to constant or even declining star-formation rates while the lower-mass galaxies are still rapidly increasing their stellar masses.

    \item The gas depletion times are quite large, about $\tau_\mathrm{dep}\gtrsim 1~\mathrm{Gyr}$, in low gas surface density environments ($\Sigma_\mathrm{HI+H_2} \lesssim 300~\Msun~\mathrm{pc}^{-2}$), but decreases to about $0.1~\mathrm{Gyr}$ at higher surface densities. This is consistent with the two different Kennicutt--Schmidt relations observed for spiral and starbursting galaxies in the local Universe.

    \item The fiducial \thzoom model includes an extra feedback channel designed to mimic the effects of some key physical processes that are missing in the current simulations and/or to compensate for the inability to resolve the physics already incorporated. This feedback, called Early Stellar Feedback (or ESF), injects momentum into adjacent gas cells for $5~$Myr immediately after a new star particle forms and is essential for producing accurate stellar masses, star-formation rates, and gradually rising, flat rotation curves in high-redshift galaxies.

    \item The ISM model is able to resolve the multi-phase nature of gas in galaxies. The amount of low-density, low-temperature gas is greatly enhanced when the widely used spatially constant UV background is replaced with a patchy radiation field taken from the parent \thesanone simulation.

    \item The gas phase metallicities and the dust-to-gas ratios of the simulated galaxies are generally in agreement with the observational estimates and  expectations from semi-analytic models, although building up significant dust in high-metallicity environments remains a challenge. The radiation-dust coupling gives rise to highly time-varying dust temperatures that seem to correlate with the rise and fall of star formation in the galaxy.

    \item  Using a variation of the method outlined in \citet{Ma2018} and \citet{Sun2023}, we are able to estimate the statistical properties of simulated galaxies like the galaxy stellar mass function (GSMF) and the UV luminosity function (UVLF). The simulated GSMF generally matches the observed values at the high-mass end ($M_\star > 10^9~\Msun$) but seems to overproduce the number density of low-mass galaxies at low redshift ($z\lesssim5.5$).

    \item The simulated UVLFs generally lie above the observed values if the fiducial dust extinction model is used. However, moving to an empirically derived dust attenuation model outlined in \citet{Bouwens2016} lowers the UV luminosities enough to match the observed estimates.

    \item The SFRD derived by integrating the UVLF beyond $M_\mathrm{UV}<-17$ matches the observational estimates from $z=3-14$. This strong agreement, along with a good match with the bright end of the UVLF, indicates that simulations effectively reproduce the star-formation rates of high-redshift galaxies observed by \jwst.
\end{enumerate}

Overall, we have shown that high-resolution coupled radiation hydrodynamic and galaxy formation simulations have the ability to robustly model the resolved properties of high-redshift galaxies. In accompanying papers, we exploit the \thzoom simulations to investigate various properties of these early galaxies like the stochasticity of star formation, statistics of LyC and \lya escape, dust and metal enrichment of the high-redshift ISM and CGM, and galaxy size evolution. We therefore hope to significantly advance our understanding of high-redshift structure formation and reionization using the \thzoom simulations.

\section*{Acknowledgments}

We thank the anonymous referees for constructive and insightful comments. The authors gratefully acknowledge the Gauss Centre for Supercomputing e.V. (\url{www.gauss-centre.eu}) for funding this project by providing computing time on the GCS Supercomputer SuperMUC-NG at Leibniz Supercomputing Centre (\url{www.lrz.de}), under project pn29we. RK acknowledges support of the Natural Sciences and Engineering Research Council of Canada (NSERC) through a Discovery Grant and a Discovery Launch Supplement (funding reference numbers RGPIN-2024-06222 and DGECR-2024-00144) and York University's Global Research Excellence Initiative. EG is grateful to the Canon Foundation Europe and the Osaka University for their support through the Canon Fellowship. Support for OZ was provided by Harvard University through the Institute for Theory and Computation Fellowship. WM thanks the Science and Technology Facilities Council (STFC) Center for Doctoral Training (CDT) in Data intensive Science at the University of Cambridge (STFC grant number 2742968) for a PhD studentship. XS acknowledges the support from the National Aeronautics and Space Administration (NASA) grant JWST-AR-04814.  LH and VS acknowledge support from
the Simons Foundation through the ``Learning the Universe"
initiative.

\section*{Data Availability}
All simulation data, including snapshots, group, and subhalo catalogs and merger trees will be made publicly available in the near future. Data will be distributed via \url{www.thesan-project.com}. Before the public data release, data underlying this article will be shared on reasonable request to the corresponding author(s).

\bibliographystyle{mnras}

\bibliography{bibliography}

\begin{appendix}

\section{Fits for stellar mass and UV luminosity functions}
\label{ap:ap1}

For easy comparison to observational data and other numerical simulations and semi-analytical models, we provide Schechter function fits \citep{S1976} for the estimated galaxy stellar mass and UV luminosity functions. Specifically, these are fits to the data plotted in Figures~\ref{fig:gsmf} and \ref{fig:uvlf}. The functional form of the Schechter function for galaxy masses is
\begin{equation}
        \Phi(M) = \mathrm{ln}(10) \,\Phi^\star \frac{\left[10^{(\mathrm{log}M - \mathrm{log}M^\star)}\right]^{\alpha+1}}{e^{10^{(\mathrm{log}M - \mathrm{log}M^\star)}}} \, ,
\end{equation}
where $M$ is the stellar mass of the galaxy (in units of $\Msun$), $\alpha$ is the
low-mass slope, $M^\star$ is the stellar mass above which the mass function
cuts off exponentially, and $\Phi^\star$ is the overall normalization. The values of these fits from $z=4-9$ are listed in Table~\ref{table:gsmf}. We find that beyond $z>9$, it is difficult to find optimal fits to the data because the \thzoom simulation suite does not contain an adequate number of high-mass galaxies in these early epochs. 
 \begin{table}
	\centering
	\caption{\textup{Schechter function fits to the galaxy stellar mass function (GSMF).}}
	\label{table:gsmf}
 \small\addtolength{\tabcolsep}{-1.0pt}
	\begin{tabular}{lccccccccc} 
		\hline
        \vspace{-0.25cm}\\
		$z$ & log($\Phi^\star$) & log(M$^\star$) & $\alpha$\vspace{0.05cm}\\  
		& [cMpc$^{-3}$ dex$^{-1}$] & [$\Msun$] & \vspace{0.05cm}\\
		\hline
        \vspace{-0.2cm}\\
		4 & $-$3.04 $\pm$ 0.18 & 9.84 $\pm$ 0.09 & $-$1.78 $\pm$ 0.03 \vspace{0.05cm}\\
        5 & $-$3.02 $\pm$ 0.30 & 9.56 $\pm$ 0.23 & $-$1.83 $\pm$ 0.03\vspace{0.05cm}\\
        6 & $-$2.98 $\pm$ 0.20 & 9.20 $\pm$ 0.11 & $-$1.88 $\pm$ 0.03\vspace{0.05cm}\\
        7 & $-$3.10 $\pm$ 0.17 & 9.03 $\pm$ 0.09 & $-$1.92 $\pm$ 0.03\vspace{0.05cm}\\
        8 & $-$3.25 $\pm$ 0.21 & 8.78 $\pm$ 0.11 & $-$1.98 $\pm$ 0.04\vspace{0.05cm}\\
        9 & $-$3.13 $\pm$ 0.24 & 8.39 $\pm$ 0.12 & $-$1.99 $\pm$ 0.05\vspace{0.05cm}\\
		\hline
	\end{tabular}
\end{table}

The functional form of the Schechter function for UV luminosities is given by
\begin{equation}
        \Phi(M_\mathrm{UV}) = \frac{0.4~\mathrm{ln}(10)~\Phi_\mathrm{UV}^\star}{10^{0.4(M_\mathrm{UV}-M_\mathrm{UV}^\star)(\alpha+1)}} e^{-10^{-0.4(M_\mathrm{UV}-M_\mathrm{UV}^\star)}} \, ,
\end{equation}
where $M_\mathrm{UV}$ is the rest frame UV magnitude of the galaxy, $\alpha$ is the
low-mass slope, $M_\mathrm{UV}^\star$ is the magnitude above which the luminosity function
cuts off exponentially, and $\Phi^\star_\mathrm{UV}$ is the overall normalization. The values of these fits from $z=4-14$ are listed in Table~\ref{table:uvlf}. It must be noted that $M_\mathrm{UV}^\star$ starts to reduce quite considerably as we get to higher redshifts ($z\gtrsim10$). This is mainly due to the fact that the \thzoom simulation suite lacks an adequate number of luminous galaxies to properly probe this high-luminosity end. 
 \begin{table}
	\centering
	\caption{\textup{Schechter function fits to the UV luminosity function (UVLF).}}
	\label{table:uvlf}
 \small\addtolength{\tabcolsep}{-1.0pt}
	\begin{tabular}{lccccccccc} 
		\hline
        \vspace{-0.25cm}\\
		$z$ & log($\Phi_\mathrm{UV}^\star$) & log($M_\mathrm{UV}^\star$) & $\alpha$\vspace{0.05cm}\\  
		& [cMpc$^{-3}$ mag$^{-1}$] &  & \vspace{0.05cm}\\
		\hline
        \vspace{-0.2cm}\\
		4 & $-$2.38 $\pm$ 0.12 & $-$20.68 $\pm$ 0.13 & $-$1.65 $\pm$ 0.02 \vspace{0.05cm}\\
        5 & $-$2.36 $\pm$ 0.17 & $-$20.53 $\pm$ 0.27 & $-$1.71 $\pm$ 0.03\vspace{0.05cm}\\
        6 & $-$2.47 $\pm$ 0.21 & $-$20.44 $\pm$ 0.31 & $-$1.77 $\pm$ 0.04\vspace{0.05cm}\\
        7 & $-$2.42 $\pm$ 0.30 & $-$20.13 $\pm$ 0.42 & $-$1.78 $\pm$ 0.05\vspace{0.05cm}\\
        8 & $-$2.44 $\pm$ 0.26 & $-$19.58 $\pm$ 0.33 & $-$1.83 $\pm$ 0.05\vspace{0.05cm}\\
        9 & $-$2.51 $\pm$ 0.28 & $-$19.29 $\pm$ 0.30 & $-$1.84 $\pm$ 0.06\vspace{0.05cm}\\
        10 & $-$2.33 $\pm$ 0.30 & $-$18.61 $\pm$ 0.28 & $-$1.84 $\pm$ 0.07\vspace{0.05cm}\\
        11 & $-$2.37 $\pm$ 0.40 & $-$18.23 $\pm$ 0.50 & $-$1.80 $\pm$ 0.10\vspace{0.05cm}\\
        12 & $-$2.21 $\pm$ 0.35 & $-$17.26 $\pm$ 0.43 & $-$1.88 $\pm$ 0.09\vspace{0.05cm}\\
        13 & $-$2.94 $\pm$ 0.60 & $-$17.92 $\pm$ 0.69 & $-$1.97 $\pm$ 0.16\vspace{0.05cm}\\
        14 & $-$3.04 $\pm$ 0.54 & $-$17.55 $\pm$ 0.70 & $-$2.07 $\pm$ 0.13\vspace{0.05cm}\\
		\hline
	\end{tabular}
\end{table}



\end{appendix}

\end{document}